\documentclass[aps,prb,floatfix,twocolumn,preprintnumbers,amsmath,amssymb,groupedaddress,showpacs,showkeys,10pt]{revtex4}

\usepackage{graphicx}  
\usepackage{dcolumn}   
\usepackage{bm}        
\usepackage{color}     
\usepackage{amsmath,amssymb}
\usepackage{slashed}

\newcommand{\llangle}{\langle\hspace{-0.4mm}\langle}
\newcommand{\rrangle}{\rangle\hspace{-0.4mm}\rangle}
\newcommand{\ua}{\uparrow}
\newcommand{\da}{\downarrow}
\newcommand{\brangle}{\bigr\rangle}
\newcommand{\blangle}{\bigl\langle}
\newcommand{\?}{\bigr|}
\newcommand{\lI}{\lambda_{\rm{I}}}
\newcommand{\lR}{\lambda_{\rm{R}}}
\newcommand{\lH}{\lambda_{\rm{PIA}}}
\newcommand{\lIA}{\lambda_{\rm{I}}^{\rm{A}}}
\newcommand{\lIB}{\lambda_{\rm{I}}^{\rm{B}}}
\newcommand{\lHA}{\lambda_{\rm{PIA}}^{\rm{A}}}
\newcommand{\lHB}{\lambda_{\rm{PIA}}^{\rm{B}}}

\newcommand{\LR}{\Lambda_{\rm{R}}}

\newcommand{\LIA}{\Lambda_{\rm{I}}^{\rm{A}}}
\newcommand{\LIB}{\Lambda_{\rm{I}}^{\rm{B}}}

\newcommand{\s}{\sigma}
\newcommand{\Sp}{\Sigma}
\newcommand{\syz}{\Sigma_{v}^{yz}}
\newcommand{\sxz}{\Sigma_{d}^{xz}}
\newcommand{\sxy}{\Sigma_{h}^{xy}}
\newcommand{\Rz}{\mathcal{R}^{\hat{z}}}

\newcommand{\rotvi}{\mathcal{R}_{\frac{\pi}{3}}^{\hat{z}}}

\newcommand{\Hh}{\hat{H}_{\mathrm{so}}}
\newcommand{\ch}{\checkmark}

\newcommand{\La}{\Lambda}

\newcommand\ess{s\kern-.17em}

\begin{document}

\title{Model spin-orbit coupling Hamiltonians for graphene systems}
\author{D. Kochan, S. Irmer and J. Fabian}
\affiliation{Institute for Theoretical Physics, University of Regensburg, 93040 Regensburg, Germany\\
 }
 \begin{abstract}
We present a detailed theoretical study of effective spin-orbit coupling (SOC) Hamiltonians for graphene based systems, covering global effects such as proximity to substrates and local SOC effects resulting, for example, from dilute adsorbate functionalization.
Our approach combines group theory and tight-binding descriptions. We consider structures with global point group symmetries $D_{6h}$, $D_{3d}$, $D_{3h}$, $C_{6v}$, and $C_{3v}$ that represent, for example, pristine graphene, graphene mini-ripple, planar boron-nitride, graphene on a substrate and free standing graphone, respectively.
The presence of certain spin-orbit coupling parameters is correlated with the absence of the specific point group symmetries. Especially in the case of $C_{6v}$---graphene on a substrate, or transverse electric field---we point out the presence of a third SOC parameter, besides the conventional intrinsic and Rashba contributions, thus far neglected in literature. For all global structures we provide effective SOC Hamiltonians both in the local atomic and Bloch forms.
Dilute adsorbate coverage results in the local point group symmetries $C_{6v}$, $C_{3v}$, and $C_{2v}$ which represent the stable adsorption at hollow, top and bridge positions, respectively. For each configuration we provide effective SOC Hamiltonians in the atomic orbital basis that respect local symmetries. In addition to giving specific analytic expressions for model SOC Hamiltonians, we also present general (no-go) arguments about the absence of certain SOC terms.


 \end{abstract}

 \date{\today}
\maketitle

\section{Introduction}

The ability to synthesize, manipulate, and functionalize 2d materials is an ultimate milestone in technological development and current fundamental research, including spintronics.\cite{Zutic2004:RMP,Fabian2007:APS} One of the major challenges is controlling, engineering, and harvesting spin degrees of freedom for faster data processing, storage, etc.
Graphene seems to be a promising material\cite{Han:Nat.Nano2014} for such applications due to its high bipolar mobility\cite{Novoselov666}, chemical and mechanical\cite{Lee385} stability, `relativistic' band structure\cite{Wallace:PR1947} with chiral electrons that are highly insensitive to backscattering,\cite{Katsnelson:NatPhys2006,Gorbachev:NanoLett2008} and, importantly for spintronics, weak intrinsic spin-orbit coupling (SOC).\cite{Gmitra2009:PRB} The latter was theoretically estimated\cite{Huertas2006,Ertler2009:PRB,Dora2010:EPL,Jeong2011:PRB,Dugaev2011:PRB} to yield long spin lifetimes---orders of microseconds---enough for harvesting electron spins as `carriers of information'. However, experiments carried out on graphene devices of the first generation gave spin lifetimes three order of magnitudes smaller.~\cite{Tombros2007:N, Tombros2008:PRL, Pi2010:PRL, Han2011:PRL, Avsar2011:NanoLett, Jo2011:PRB, Mani2012:NC}
This vast discrepancy can be reliably explained assuming a small amount (orders of ppm) of resonant magnetic scatters\cite{Kochan2014:PRL,Kochan2015resonant,Kochan2015:PRL} like for example hydrogen atoms\cite{Wehling2010:PRL,Yazyev2008:PRL} or vacancies.\cite{Pereira2006:PRL,Yazyev2008:PRL}
Related theoretical studies\cite{Soriano2015:2DMaterials,Thomsen2015:PRB,Wilhem2015:PRB} confirmed that magnetic moments, indeed, strongly affect spin dynamics and can cause the ultra-fast spin relaxation. A recent experiment of the Valenzuela group [\onlinecite{Reas2016:NatComm}], analyzing graphene's spin-lifetime anisotropy, supports that view and convincingly rules out SOC as a determining factor of the fast spin relaxation.

On the other hand, enhancing SOC in graphene is desirable as well. Indeed, graphene with strong intrinsic SOC is predicted to host the quantum spin Hall phase.\cite{Kane2005:PRL} Therefore, one of the current technological and theoretical challenges is to tailor the strength of SOC of graphene in a controllable manner. In fact, SOC can be significantly enhanced either by chemical functionalization---coating of graphene with light\cite{CastroNeto2009:PRL,Gmitra2013:PRL,Irmer2015:PRB,Zollner2016:PRB,Frank2016:Cu,Balakrishnan2013:NP,Balakrishnan2014:NatCom} or heavy\cite{Weeks2011:PRX,Ma2012:Carbon,Hu2012:PRL,Acosta2014:PRB} adatoms accompanied by band gap opening---or by a variety of proximity effects resulting from substrates or due to scaffolding of different 2d materials\cite{GeimGrigorieva:Nature2013}.
Tangible examples are CVD graphene grown on Cu and Ni substrates\cite{Frank2016:PRB,Varykhalov2008:PRL,Shikin2013:NJP}, or graphene placed on top of transition metal dichalcogenides\cite{KaloniAPL:2014,Avsar2014:NComm,Morpurgo2015:NComm,Gmitra2016:PRB(2)}.

To further examine SOC effects in functionalized graphene and also design device properties, one needs an effective model that allows
reliable simulations of the spin and charge transport characteristics.\cite{StauberSchliemann2009:NJP,MingHaoLiu2012:PRB,Scholz2012:PRB,Ferreira2014:PRL,Pachoud2014:PRB,VanTuan2014:NP,Bundesmann2015:PRB,Soriano2015:2DMaterials,MingHaoLiu2015:PRL,Garcia2016:2DMaterials,Cummings2016:PRL,ChangNikolic2014:NanoLett,VanTuan2016:PRL1,VanTuan2016:PRL2}
In this paper we present a detailed symmetry analysis focusing on effective SOC Hamiltonians in a way that is complementary to Refs.~[\onlinecite{Weeks2011:PRX,Pachoud2014:PRB,Brey2015:PRB}]. Our findings remain valid for any hexagonal (graphene-like) structure possessing $\pi$-orbitals and are easily transferable to other systems. The primary aim of this manuscript is to lift the curtain and show practically how to derive the corresponding SOC Hamiltonians from the given pools of global or local symmetries.

We discuss two cases: global SOC Hamiltonians that represent proximity induced phenomena or periodically functionalized structures, and local SOC Hamiltonians that govern spin dynamics in the vicinity of adsorbates. Starting with pristine graphene, we step-by-step reduce the number of global symmetries approaching structures such as graphene mini-ripple, staggered graphene, planar boron-nitride, silicene, graphene on a substrate, graphone, etc. For each representative case, which is classified by the associated subgroup of the full hexagonal group, we derive an effective SOC Hamiltonian in real and reciprocal space, respectively. Our analysis therefore covers also quasi-momenta that are not necessarily constrained to the vicinity of Dirac points.

In the case of local impurities we focus on the reduction of local symmetries up to a certain spatial extent from the adsorbate. The three representative adsorption positions are hollow, top, and bridge and we provide here the local SOC Hamiltonians in real space.
Group arguments allow us to link the presence or absence of certain symmetries to various spin-orbit couplings that emerge in the effective SOC Hamiltonian. For example, in the global case corresponding to point group $C_{6v}$---graphene in a transverse electric field or deposited on a substrate---we highlight the presence of a SOC term that have not yet been considered. It appears along with the conventional intrinsic and Rashba couplings and is related to the absence of the principal mirror plane in the structure.

The paper is organized as follows. After recapitulating the basic group theory related with the full hexagonal system and its application to SOC matrix elements in Sec.~\ref{sec:GT_SOC}, we discuss separately translational invariant systems, Sec.~\ref{sec:trans_inv}, and systems lacking that invariance (local adsorbates), Sec.~\ref{sec:local}.
In subsections of \ref{sec:trans_inv}, we cover in detail SOC in pristine graphene, point group $D_{6h}$, and effective SOC Hamiltonians in systems that are characterized by one of its subgroups: $D_{3d}$, $D_{3h}$, $C_{6v}$ and $C_{3v}$.
Section \ref{sec:local} is devoted to local SOC Hamiltonians for the three stable adsorption positions---hollow, top and bridge, respectively.
Summary and final remarks are provided in Sec.~\ref{sec:conclude}.

\section{Group theory and SOC - preliminaries}\label{sec:GT_SOC}

A convenient approach how the group theory enters effective model building is a decomposition of the Hamiltonian matrix elements associated with the problem into irreducible representations (irreps). Those are well known and standardly tabulated for all crystallographic point groups~\cite{KDWS1963:MIT,Dresselhaus2008}.
Considering spin and spin-orbit interaction the irrep analysis around the high symmetry points in the Brillouin zone becomes more involved. This is because the associated double (also called spinor) group representations should be appropriately taken into account; the case of graphite is exhaustively discussed in the thesis of Slonczewski~[\onlinecite{Slonczewski1955:PhD-thesis}].
For a general overview and connection with the theory of group invariants, see the book of Bir and Pikus~[\onlinecite{BirPikus1974}], or Winkler~[\onlinecite{Winkler2003}].

Another possibility how to derive an effective SOC Hamiltonian is to employ the multi-orbital tight-binding approach.\cite{Konschuh2010:PRB,Konschuh2011:PhDThesis,Liu2011:PRB,Geissler2013:NJP}
The group symmetry analysis on the orbital level is straightforward and well described by the Koster-Slater two-center approximation\cite{KosterSlater:PR1954} and, consequently, SOC enters as the intra-atomic LS-interaction $\xi_\ell\,\hat{\mathbf{L}}\cdot\hat{\mathbf{S}}$. The resulting multi-orbital tight-binding Hamiltonian is then down-folded by means
of the L\"owdin projection\cite{LowdinJChemPhys:1951} to the states of interest---mostly the low energy states with respect to the Fermi level.

As an alternative to the invariant expansion and the multi-orbital tight-binding method with the L\"owdin projection, we present here an effective tight-binding approach that employs symmetries of local atomic orbitals.
We focus particularly on hexagonal lattice structures assuming the low energy physics near the Fermi level can be approximately well described by $\pi$-orbitals, i.e.~carbon $2p_z$ orbitals, or atomic orbitals $n,\ell\neq 0,m_\ell=0$.
For simplicity we consider that each nodal atomic site $m$ contains one effective $\pi$-orbital state, $|X_m\rangle\equiv c^\dagger_{m}|0\rangle$.
When it is necessary to specify the sublattice $X$, we explicitly write $|A_m\rangle$ and $|B_m\rangle$, for the two atomic sites in a hexagonal lattice. Including also electron spin, $\s=\{\ua,\da\}\equiv\{+1,-1\}$, the effective one-particle Hilbert space is spanned by states $|X_m\,\s\rangle\equiv c^\dagger_{m,\s}|0\rangle$.

\begin{figure}
\includegraphics[width=\columnwidth]{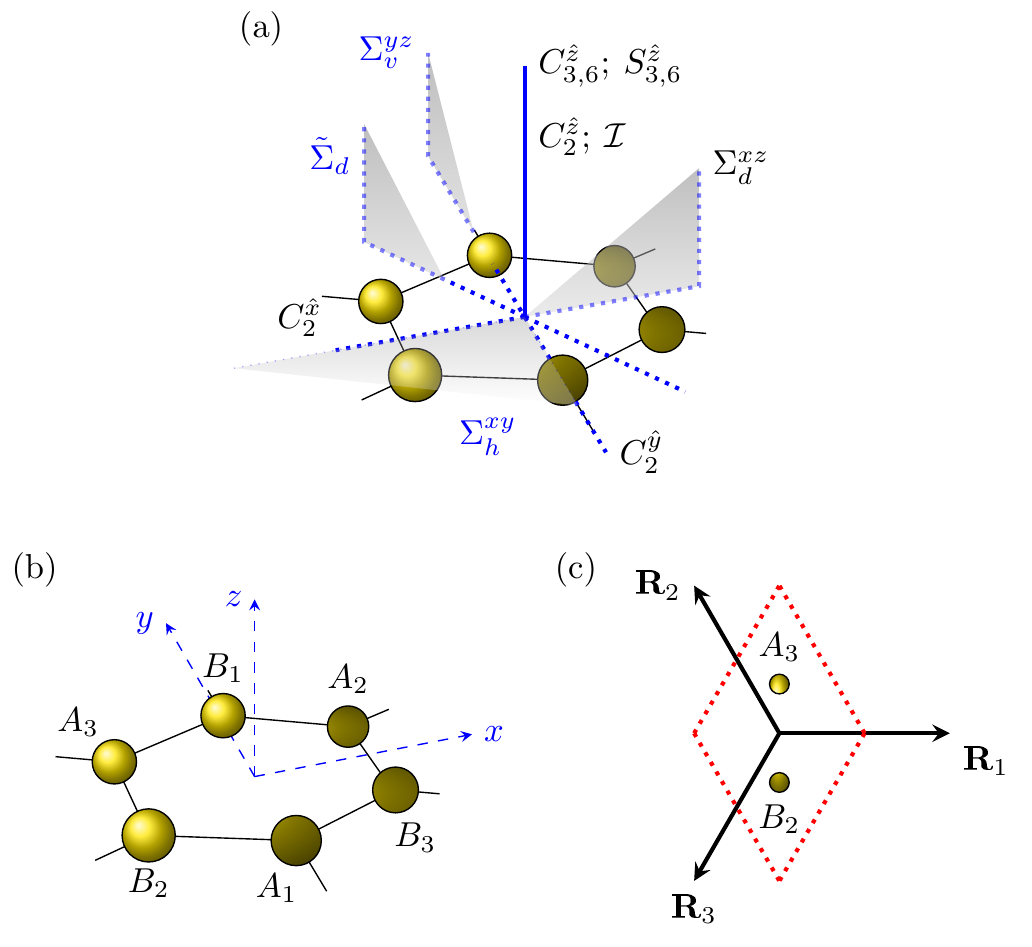}
\caption{(Color online) Panel (a) shows selected symmetry operations of point group
$D_{6h}$. Combining the point group generators (blue symbols)---horizontal $\sxy$, vertical $\syz$, and dihedral $\tilde{\Sp}_d$ reflections, one can built all the remaining group elements (black symbols). Lower panels depict (b) sites' labeling convention and axes orientation, and (c) graphene unit cell together with the Bravais lattice vectors $\textbf{R}_\alpha$ ($\alpha=1,2,3$).
\label{fig:hex_sitelabel_coordinate}}
\end{figure}

The structural point group of an ideal hexagonal lattice is the symmetry group $D_{6h}$---in international crystallographic notation group $6/\emph{mmm}$. It contains 24 group elements which can be expressed in terms of four group generators: identity $E$ and reflections $\Sp_h\equiv\sxy,\Sp_v\equiv\syz,\tilde{\Sp}_d$, for visualization see Fig.~\ref{fig:hex_sitelabel_coordinate}.
Subscripts $h$, $v$, and $d$ stand for the horizontal ($xy$-plane), vertical ($yz$-plane), and dihedral (rotated $xz$-plane) reflections, respectively. When it is convenient to emphasize the reflection planes explicitly, we employ the superscripts $xz$, $yz$, and $xy$. Similarly, to specify the axis determining a spatial rotation we use hat superscripts, such as $\hat{x}$, $\hat{y}$, and $\hat{z}$. The remaining elements of $D_{6h}$ are 6-fold and 3-fold rotations, $C_6^{\hat{z}}=\tilde{\Sp}_d\circ\Sp_v$ and $C_3^{\hat{z}}=C_6^{\hat{z}}\circ C_6^{\hat{z}}$, the $xz$-dihedral reflection $\Sp_d\equiv\sxz=C_3^{\hat{z}}\circ\tilde{\Sp}_d$, the space-inversion $\mathcal{I}=\Sp_h\circ\Sp_d\circ\Sp_v$, the improper rotations
$S_{6}^{\hat{z}}=\Sp_h\circ C^{\hat{z}}_{6}$, $S_{3}^{\hat{z}}=\Sp_h\circ C^{\hat{z}}_{3}$, and the 2-fold rotations $C_2^{\hat{x}}=\Sp_h\circ\Sp_d$, $C_2^{\hat{y}}=\Sp_h\circ\Sp_v$, and $C_2^{\hat{z}}=\Sp_v\circ\Sp_d$, see Fig.~\ref{fig:hex_sitelabel_coordinate}.

To construct an invariant SOC Hamiltonian it is necessary to know how the one-particle basis states $|X_m\,\s\rangle$ transform under the active action of $D_{6h}$ including the time-reversal symmetry $\mathcal{T}$.
While we are not dealing with the double group irreps
it is enough to focus on the action of selected group elements: rotation ${\cal R}^{\hat{z}}_\Phi$ by an angle $\Phi$, the horizontal, vertical, and dihedral reflections $\Sp_h$, $\Sp_v$, and $\Sp_d$, respectively, the time-reversal $\cal T$, and for completeness also the space-inversion ${\cal I}$ and the translation $T_{\vec{a}}$ by a lattice vector $\vec{a}$:
\begin{subequations}\label{eq:realspace_trafo}
\begin{align}
|X_m\,\s\brangle\ &\xrightarrow{\,\mathcal{R}_\Phi\,}\ e^{-i\sigma\frac{\Phi}{2}}\,|X_{\mathcal{R}_\Phi(m)}\,\s\brangle\label{eq:r_rotation}\,,\\
|X_m\,\s\brangle\ &\xrightarrow{\,\sxy\,}\ i(-1)^{\frac{1+\sigma}{2}}\,|X_m\,\s\brangle\label{eq:r_reflection_h}\,,\\
|X_m\,\s\brangle\ &\xrightarrow{\,\syz\,}\ i\,|X_{\syz(m)}\,(-\s)\brangle\label{eq:r_reflection_yz}\,,\\
|X_m\,\s\brangle\ &\xrightarrow{\,\sxz\,}\ (-1)^{\frac{1+\sigma}{2}}\,|X_{\sxz(m)}\,(-\s)\brangle\label{eq:r_reflection_xz}\,,\\
|X_m\,\s\brangle\ &\xrightarrow{\ \,\mathcal{T}\,\ }\ (-1)^{\frac{1-\sigma}{2}}\,|X_m\,(-\s)\brangle\label{eq:r_timereversal}\,,\\
|X_m\,\s\brangle\ &\xrightarrow{\ \ \mathcal{I}\,\ }\ -|X_{\mathcal{I}(m)}\,\s\brangle\label{eq:r_spaceinversion}\,,\\
|X_m\,\s\brangle\ &\xrightarrow{\ \,T_{\vec{a}}\,\ }\ |X_{\,m+\vec{a}}\,\s\brangle\label{eq:r_translation}\,.
\end{align}
\end{subequations}
The action of the remaining $D_{6h}$ elements follow immediately from the relations to the group generators. The action of $\cal T$ affects only the spin component of $|X_m\,\s\rangle=|X_m\rangle\otimes|\s\rangle$ since, by convention, our orbital $\pi$-states $|X_m\rangle$ are real-valued wave-functions.

An electron moving in an effective crystal field potential $V$ is affected by SOC interaction that is represented by Hamiltonian,
\begin{equation}\label{eq:micro_soc}
\hat{H}_{\rm so}=\frac{\hbar}{4m_{\rm e}^2{\rm c}^2}\bigl({\boldsymbol{\nabla}} V\times\hat{\mathbf{p}}\bigr)\cdot\hat{\boldsymbol{s}}\,.
\end{equation}
Here, $m_{\rm e}$ is the vacuum rest mass of the electron, $\rm c$ the speed of light, $\hat{\mathbf{p}}$ stands for the momentum operator, and
$\hat{\boldsymbol{s}}=(\hat{s}_x,\hat{s}_y,\hat{s}_z)$ represents the array of Pauli matrices acting on spin degrees of freedom. In reality we do not know the crystal field and so $\hat{H}_{\rm so}$ exactly, but knowing the pool of symmetries preserving $V$, and hence $\hat{H}_{\rm so}$, we can uniquely detect which matrix elements
$\langle X_m\,\s|\hat{H}_{\rm so}|X_n\,\s^\prime\rangle$ are non-zero and thus important.
If $\mathcal{S}$ is a system's symmetry---precisely, its {\it{unitary}} representation---then $\mathcal{S}\,\hat{H}_{\rm so}=\hat{H}_{\rm so}\,\mathcal{S}$ and
\begin{align}
\blangle \mathcal{S}[X_m\,\s]\,|\,\hat{H}_{\rm so}\,|\,\mathcal{S}[X_n\,\s']\brangle &=
\blangle \mathcal{S}[X_m\,\s]\,|\,\mathcal{S}[\hat{H}_{\rm so}\,X_n\,\s']\brangle \nonumber\\
&=\blangle X_m\,\s\,|\,\hat{H}_{\rm so}\,|\, X_n\,\s'\brangle\label{Eq.:Unitary-Action}
\end{align}
for any two one-particle states $|X_m\,\s\rangle$ and $|X_n\,\s'\rangle$. In an analogous way we get for the {\it{anti-unitary}} time-reversal symmetry,
$\mathcal{T}\,\Hh=\Hh\,\mathcal{T}$, and self-adjoint $\Hh$
\begin{align}
\blangle \mathcal{T}[X_m\,\s]\,|\,\Hh\,|\,\mathcal{T}[X_n\,\s']\brangle &=
\blangle \mathcal{T}[X_m\,\s]\,|\,\mathcal{T}[\Hh\,X_n\,\s']\brangle \nonumber\\
&=\overline{\blangle X_m\,\s\,|\,\Hh\,|\, X_n\,\s'\brangle}\nonumber\\
&=\blangle X_n\,\s'\,|\,\Hh\,|\,X_m\,\s\brangle\,.\label{Eq.:AntiUnitary-Action}
\end{align}
This gives us a practical relation connecting SOC matrix elements with opposite spin projections:
\begin{align}
&\blangle X_m\,\s\?\Hh\?X_n\,\s' \brangle=\nonumber\\
&\overset{(\ref{eq:r_timereversal})}{=}\blangle (-1)^{-\frac{1-\s}{2}}\mathcal{T}[X_m(-\s)]\?\Hh\? (-1)^{-\frac{1-\s'}{2}}\mathcal{T}[X_n(-\s')] \brangle\nonumber\\
&=-(-1)^{\frac{\s+\s'}{2}}\,\blangle \mathcal{T}[X_m(-\s)]\?\Hh\? \mathcal{T}[X_n(-\s')] \brangle\nonumber\\
&\overset{(\ref{Eq.:AntiUnitary-Action})}{=}-(-1)^{\frac{\s+\s'}{2}}\, \blangle X_n(-\s')\?\Hh\?X_m(-\s) \brangle\,.\label{Eq.:opposite spin SOC elements}
\end{align}

In practice, we focus only on the on-site, nearest neighbors, and the next nearest neighbors SOC mediated hoppings $\langle X_m\,\s|\hat{H}_{\rm so}|X_n\,\s^\prime\rangle$. This is sufficient because the orbital overlaps modulated by $\boldsymbol{\nabla} V$---dominant near the atomic cores---decay rather fast with increasing distance. Therefore we focus on SOC hoppings $\langle X_m\,\s|\hat{H}_{\rm so}|X_n\,\s^\prime\rangle$ inside one particular elementary cell of the hexagonal lattice, see Fig.~\ref{fig:hex_sitelabel_coordinate}. All other spin-resolved hoppings can be expressed by applying translations, rotations, reflections, or time-reversal, see Eqs.~(\ref{eq:realspace_trafo}).

In what follows we show how time-reversal symmetry and self-adjointness of $\hat{H}_{\rm so}$ restrict $\langle X_m\,\s|\hat{H}_{\rm so}|X_n\,\s^\prime\rangle$. Particulary, we argue that the spin-conserving hoppings $\langle X_m\,\s|\Hh|X_n\,\s\rangle$ are purely imaginary, and the on-site SOC resolved hoppings
$\langle X_m\,\s|\hat{H}_{\rm so}|X_m\,\s\rangle$ and $\langle X_m\,\s|\hat{H}_{\rm so}|X_m(-\s)\rangle$ vanish.
First, note that the SOC Hamiltonian $\hat{H}_{\rm so}$, Eq.~(\ref{eq:micro_soc}), can be recast into the form,
\begin{equation}\label{eq:micro_soc_via_L-operators}
\hat{H}_{\rm so}=\hat{\mathcal{L}}_{+}\hat{s}_{-}+\hat{\mathcal{L}}_{-}\hat{s}_{+}+\hat{\mathcal{L}}_{z}\hat{s}_{z}\,,
\end{equation}
where $\hat{s}_{\pm}=\tfrac{1}{2}(s_x\pm is_y)$ are spin raising and lowering operators (without $\tfrac{\hbar}{2}$) and $\hat{\mathcal{L}}$'s act solely on the orbital part of the wave-function. It follows from the hermiticity of $\Hh$ that $\hat{\mathcal{L}}_{-}^\dagger=\hat{\mathcal{L}}_{+}$ and $\hat{\mathcal{L}}_z$ is self-adjoint. Also $\mathcal{L}$'s transform under the space and time reversal symmetries equally as the standard angular momentum operators. However, for a general crystal field potential $V$ they do not obey the usual SU(2)-commutation relations. Directly from Eq.~(\ref{eq:micro_soc_via_L-operators}) we have
\begin{equation}\label{Eq.:pure_imaginary2}
\blangle X_m\,\s\?\Hh\?X_n\,\s\brangle=-\blangle X_m(-\s)\?\Hh\?X_n(-\s)\brangle\,.
\end{equation}
On the other side, the time-reversal symmetry, Eq.~(\ref{Eq.:opposite spin SOC elements}), implies:
\begin{align}
\blangle X_m(-\s)\?\Hh\?X_n(-\s)\brangle &\overset{(\ref{Eq.:opposite spin SOC elements})}{=} \blangle X_n\,\s\?\Hh\? X_m\,\s\brangle\nonumber\\
&=\overline{\blangle X_m\,\s\?\Hh\?X_n\,\s\brangle}\,.
\end{align}
So comparing this and the above expression we see that
\begin{align}\label{Eq.:pure_imaginary}
&\blangle X_m\,\s\?\Hh\?X_n\,\s\brangle &  &\text{\emph{is a purely imaginary SOC}} \nonumber \\
&{} &  &\hspace{0.8cm}\text{\emph{matrix element}}
\end{align}
\emph{for any two atomic sites mediating a spin-conserving hopping}. In the special case $m=n$ the above
Eqs.~(\ref{Eq.:opposite spin SOC elements})~and~(\ref{Eq.:pure_imaginary2}) give:
\begin{align}
\blangle X_m\,\s\?\Hh\?X_m\,\s\brangle &\overset{(\ref{Eq.:opposite spin SOC elements})}{=} \blangle X_m(-\s)\?\Hh\?X_m(-\s)\brangle\nonumber\\
&\overset{(\ref{Eq.:pure_imaginary2})}{=}-\blangle X_m\,\s\?\Hh\?X_m\,\s\brangle\,,
\end{align}
so that we have shown that \emph{the on-site spin-conserving term $\langle X_m\,\s|\Hh|X_m\,\s\rangle$ equals zero for any site $m$}.
In a similar way we get for its spin-flipping counterpart:
\begin{align}
\blangle X_m\,\s\?\Hh\?X_m(-\s)\brangle &\overset{(\ref{Eq.:opposite spin SOC elements})}{=} -\blangle X_m\,\s\?\Hh\?X_m(-\s)\brangle\,,
\end{align}
so \emph{the on-site spin-flipping matrix element $\langle X_m\,\s|\Hh|X_m(-\s)\rangle$ is
zero for any lattice site $m$}. Therefore, what matters are the nearest and next nearest neighbors SOC mediated matrix elements which
we will examine in the forthcoming sections.

\section{Translational invariant systems}\label{sec:trans_inv}
\subsection{Pristine graphene SOC Hamiltonian} \label{sec:pristine_graphene}

The spin-orbit coupling Hamiltonian based on $\pi$-states that is translational invariant and possesses the full point group symmetry $D_{6h}$ of the pristine graphene allows only one, the so called intrinsic, SOC hopping $\lI$. This was first discussed by McClure and Yafet~\cite{McClure1962} when analyzing the g-factor in a ``graphite single crystal''. Later Kane and Mele~\cite{Kane_Mele_2_2005:PRL} revisited this point when predicting the quantum spin Hall effect in graphene.
The magnitude of $\lI$ was found in the work of Gmitra et al.~[\onlinecite{Gmitra2009:PRB}], who showed that $\lI$ is too weak\cite{Gmitra2009:PRB}---about 12~$\mu$eV---to induce an experimentally detectable transition into the quantum spin Hall phase. Furthermore, Gmitra et al.~[\onlinecite{Gmitra2009:PRB}] found that $\lI$ is due to the coupling of $p_z$ and $d$ orbitals. This was supported
by multi-orbital tight-binding calculations; Konschuh et al.~[\onlinecite{Konschuh2010:PRB}] showed that the intrinsic SOC hopping $\lI$ is significantly affected by the admixture of $3d_{xz}\pm i3d_{yz}$-orbitals, the fact anticipated already by Slonczewski [\onlinecite{Slonczewski1955:PhD-thesis}].

The effective tight-binding Hamiltonian mediating the SOC interaction among $\pi$-states in graphene---or any planar hexagonal system with one $\pi$-orbital per site---reads
\begin{align}\label{Eq.:pristine-graphene-SOC-Hamiltonian}
\mathcal{H}_{D_{6h}}&=\frac{i\lI}{3\sqrt{3}}
\sum\limits_\sigma\sum\limits_{\llangle m,n\rrangle}
\nu_{m,n}^{\phantom\dagger}\bigl[\hat{s}_z\bigr]_{\sigma\sigma}\?X_m\,\s\brangle\,\blangle X_n\,\s\?\,.
\end{align}
The Hamiltonian $\mathcal{H}_{D_{6h}}$ couples next nearest neighbors (summation over $\llangle m,n\rrangle$) and allows only spin-conserving hoppings. Therefore, in accordance with Eq.~(\ref{Eq.:pure_imaginary}) the underlying coupling constant is purely imaginary. Using the configuration shown at Fig.~\ref{fig:hex_sitelabel_coordinate}, the coupling $i\lI$ can be defined as,
\begin{subequations}\label{Eq.:intr_graphene}
\begin{equation}\label{Eq.:intr_graphene1}
\frac{i\lI}{3\sqrt{3}}=\blangle A_3\ua\?\Hh\?A_2\ua \brangle\overset{(\ref{Eq.:pure_imaginary2})}{=}-\blangle A_3\da\?\Hh\?A_2\da \brangle\,.
\end{equation}
The numerical factor $1\bigl/3\sqrt{3}$ is a matter of convention; adding it here, the low energy expansion of the Bloch transform of $\mathcal{H}_{D_{6h}}$ becomes simpler. In the above formula and also below, we
identify a lattice site $m$ with a $\pi$-state $|X_m\rangle$ residing on it. Since each site hosts one $\pi$-orbital state, this assignment is unique. Moreover, since $|A_2\ua\rangle=\sxz |B_3 \da\rangle$, and $|A_3\ua\rangle=\sxz |B_2 \da\rangle$, see Fig.~\ref{fig:hex_sitelabel_coordinate}, we can write
\begin{align}\label{Eq.:intr_graphene2}
\frac{i\lI}{3\sqrt{3}}&=\blangle A_3\ua\?\Hh\?A_2\ua \brangle\overset{(\ref{eq:r_reflection_xz})}{=}\blangle \sxz[B_2\da]\?\Hh\?\sxz[B_3\da] \brangle\nonumber\\
&\overset{(\ref{Eq.:Unitary-Action})}{=}\blangle B_2\da\?\Hh\?B_3\da \brangle\overset{(\ref{Eq.:pure_imaginary2})}{=}-\blangle B_2\ua\?\Hh\?B_3\ua \brangle\,.
\end{align}
\end{subequations}
All the sublattice and spin related sign factors are captured in the prefactor term $\nu_{m,n}^{\phantom\dagger}\,[\hat{s}_z]_{\sigma\sigma}$, i.e.,
\begin{equation}\label{Eq.:intr_graphene3}
\blangle X_m\,\s\?\Hh\?X_n\,\s \brangle=\nu_{m,n}^{\phantom\dagger}\,[\hat{s}_z]_{\sigma\sigma}\,\frac{i\lI}{3\sqrt{3}}\,.
\end{equation}
There, $\nu_{m,n}=+1(-1)$, if the next nearest neighbor hopping $n\rightarrow m$ via a common neighbor on the opposite sublattice is counter clockwise (clockwise), e.g.,~for $A_2\rightarrow (B_1)\rightarrow A_3$, $\nu_{A_3A_2}=+1$, while
for $B_3\rightarrow (A_1)\rightarrow B_2$, $\nu_{B_2B_3}=-1$, see Fig.~\ref{fig:hex_sitelabel_coordinate}. The dependence on spin $\s$ is governed by $[\hat{s}_z]_{\sigma\sigma}$; as defined $[\hat{s}_z]_{\pm\pm}=\pm 1$.

To see the effect of the intrinsic SOC on the band structure we transform $\mathcal{H}_{D_{6h}}$, Eq.~(\ref{Eq.:pristine-graphene-SOC-Hamiltonian}), from the local atomic into the Bloch basis, $|X_m\,\s\rangle\mapsto |X_{\mathbf{q}}\,\s\rangle$:
\begin{equation}
\?X_{\mathbf{q}}\,\s\brangle=\frac{1}{\sqrt{N_1N_2}}\,\sum\limits_{\mathbf{R}_m}\, e^{i\mathbf{q}\cdot\mathbf{R}_m}\,\?X_m\,\s\brangle\,.
\end{equation}
Here, $X=\{A,B\}$ and $\s=\{\ua,\da\}$, dependent on the sublattice and spin degrees of freedom, respectively,
$\mathbf{q}$ is the quasi-momentum measured from the center of the hexagonal Brillouin zone ($\Gamma$ point), $N_1N_2$ is the number of graphene unit cells in the sample, and $\mathbf{R}_m$ is the lattice vector of the $m$-th cell that hosts the orbital $|X_m\,\s\rangle$.
Inserting the above unitary transformation into Eq.~(\ref{Eq.:pristine-graphene-SOC-Hamiltonian}) we transform $\mathcal{H}_{D_{6h}}$ to the Bloch form,
$\mathcal{H}_{D_{6h}}=\sum_{\mathbf{q}}\mathcal{H}_{D_{6h}}(\mathbf{q})$, where
\begin{align}\label{Eq.:pristine-graphene-SOC-Hamiltonian Bloch Form}
\mathcal{H}_{D_{6h}}(\mathbf{q})=\lI\,f_{\mathrm{I}}(\mathbf{q})\,\sum\limits_{X,\s}\,\bigl[\hat{\s}_z\bigr]_{XX}\bigl[\hat{s}_z\bigr]_{\s\s}\,
\?X_{\mathbf{q}}\,\s\brangle\blangle X_{\mathbf{q}}\,\s\?\,.
\end{align}
Here, the Pauli matrix $\hat{\s}_z$ acts in the space of sublattices---$[\hat{\s}_z]_{AA}=1=-[\hat{\s}_z]_{BB}$, and $[\hat{\s}_z]_{AB}=0=[\hat{\s}_z]_{BA}$. The intrinsic structural function $f_{\mathrm{I}}(\mathbf{q})$ reads,
\begin{equation}\label{Eq:structural-I-funcion}
f_{\mathrm{I}}(\mathbf{q})=-\frac{2}{3\sqrt{3}}\bigl\{\sin{\mathbf{q}\cdot\mathbf{R}_1}+\sin{\mathbf{q}\cdot\mathbf{R}_2}+\sin{\mathbf{q}\cdot\mathbf{R}_3}\bigr\}\,.
\end{equation}
The lattice vectors $\mathbf{R}_\alpha$ ($\alpha=1,2,3$) can be compactly expressed in terms of the Levi-Civita antisymmetric $\epsilon$-symbol and the position vectors of the lattice sites $A_1$, $A_2$, and $A_3$ as displayed at Fig.~\ref{fig:hex_sitelabel_coordinate}. Particularly,
\begin{equation}
\mathbf{R}_\alpha=\tfrac{1}{2}\epsilon_{\alpha\beta\gamma}\overrightarrow{A_\gamma A_\beta}=a_{\mathrm{L}}\bigl(\cos{\tfrac{2\pi(\alpha-1)}{3}},\sin{\tfrac{2\pi(\alpha-1)}{3}}\bigr)\,,
\end{equation}
where $a_{\mathrm{L}}$ is the lattice constant; in the case of graphene $a_{\mathrm{L}}=2.46$\,{\AA}.

On the orbital level the electronic band structure of graphene $\pi$-orbitals is well described by the standard nearest neighbor Hamiltonian,
\begin{equation}\label{Eq:pristine-graphene orbital}
\mathcal{H}_\mathrm{orb} = -t\,\sum\limits_{\s}\sum\limits_{\langle m,n\rangle}\?X_m\,\s\brangle \blangle X_n\,\s\?\,,
\end{equation}
with $t=2.6\,$eV. Transforming it to the Bloch form we arrive at $\mathcal{H}_\mathrm{orb}=\sum_{\mathbf{q}}\mathcal{H}_\mathrm{orb}(\mathbf{q})$, where,
\begin{equation}\label{Eq:pristine-graphene orbital Bloch}
\mathcal{H}_\mathrm{orb}(\mathbf{q}) = -t\,\sum\limits_{\s} f_{\mathrm{orb}}(\mathbf{q})\,\bigl[\hat{s}_0\bigr]_{\s\s} \?A_\mathbf{q}\,\s\brangle \blangle B_\mathbf{q}\,\s\?+\mathrm{hc}\,,
\end{equation}
and the orbital structural function is given by
\begin{equation}
f_{\mathrm{orb}}(\mathbf{q}) = \bigl\{1+\mathrm{e}^{i\mathbf{q}\cdot\mathbf{R}_2}+\mathrm{e}^{-i\mathbf{q}\cdot\mathbf{R}_3}\bigr\}\,;
\end{equation}
$\hat{s}_0$ is the identity matrix in spin space.

In what follows we focus on the low energy physics near the Dirac points,
\begin{equation}
\pm\mathbf{K}=\pm\frac{4\pi}{3a_{\mathrm{L}}}(1,0)\,,
\end{equation}
i.e.,~we substitute for $\mathbf{q}=\pm\mathbf{K}+\mathbf{k}$ and expand the relevant $\mathbf{q}$-dependent quantities in $\mathbf{k}$ keeping the first
non-zero term. For the above defined structural functions we particularly get,
\begin{equation}
f_{\mathrm{I}}(\pm\mathbf{K}+\mathbf{k})\simeq \pm 1\,,
\end{equation}
and
\begin{equation}
f_{\mathrm{orb}}(\pm\mathbf{K}+\mathbf{k})\simeq \frac{\sqrt{3}a_{\mathrm{L}}}{2}\,\bigl(\mp k_x - i k_y\bigr)\,,
\end{equation}
Fixing the order $\{|A_{\mathbf{q}}\ua\rangle$, $|A_{\mathbf{q}}\da\rangle,|B_{\mathbf{q}}\ua\rangle,|B_{\mathbf{q}}\da\rangle\}$ of the Bloch basis, we arrive at the effective low energy Hamiltonian in the form,
\begin{equation}\label{Eq.:Effective-D6h}
\mathcal{H}_{\mathrm{eff}}(\tau\mathbf{K}+\mathbf{k})=\hbar v_F\,\bigl(\tau k_x\hat{\s}_x-k_y\hat{\s}_y\bigr)\hat{s}_0+\tau\lI\hat{\s}_z\hat{s}_z\,.
\end{equation}
Here, $\tau\mathbf{K}=\pm\mathbf{K}$ is the shorthand for the Dirac valleys, $\hat{\s}_{x(y)}$ are Pauli matrices in the sublattice space,
and $v_F=\sqrt{3}a_{\mathrm{L}}t/2\hbar$ stands for the Fermi velocity; for example for graphene $v_F\approx 10^6\rm{m/s}$.

\begin{figure}
\includegraphics[width=0.65\columnwidth]{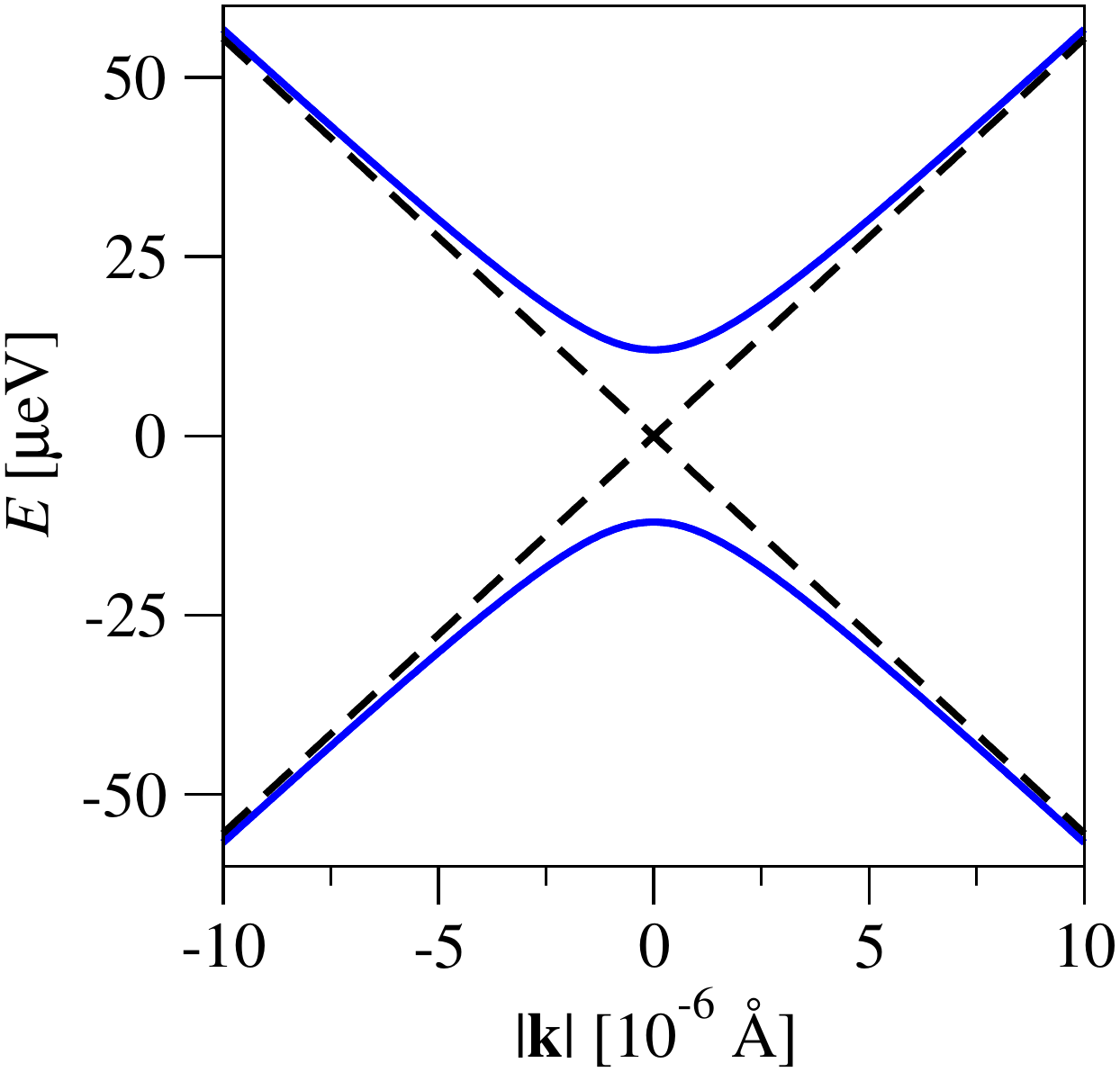}
\caption{(Color online) Electronic band structure of the typical graphene-like system in the vicinity of the Dirac point ($|\mathbf{k}|=0$) with (solid line) and without (dashed line) the intrinsic ($D_{6h}$ invariant) SOC Hamiltonian. The spectral gap of
$2\lI$ and the parabolic shape near $|\mathbf{k}|=0$ are typical imprints of the intrinsic SOC which does not spin-split the bands.
\label{fig:d6h_li12}}
\end{figure}

From the above Bloch representation we see that $\hat{\s}_0\hat{s}_z$ commutes with $\mathcal{H}_{\mathrm{eff}}$, and hence its eigenstates can be labeled by the spin $\ua$ and $\da$ projections along the $z$ spin quantization axis independently of $\mathbf{k}$. The eigenspectrum of $\mathcal{H}_{\mathrm{eff}}$, dependent on the quasimomentum $\mathbf{k}$, band index $n=+/-$\,=\,conduction/valence, and spin $\s=\{\ua,\da\}$, reads,
\begin{equation}
\varepsilon_{n,\s}(\tau\mathbf{K}+\mathbf{k})=n\sqrt{\lI^2+\hbar^2 v_F^2 (k_x^2+k_y^2)}\,.
\end{equation}
The corresponding four eigenstates get grouped into pairs, each pair comprising states with the opposite spins,~e.g.,~directly at the $\tau\mathbf{K}$ points we have two pairs $\{|A_{\tau\mathbf{K}}\ua\rangle, |B_{\tau\mathbf{K}}\da\rangle\}$ and $\{|A_{\tau\mathbf{K}}\da\rangle,|B_{\tau\mathbf{K}}\ua\rangle\}$, that are split in energy by the intrinsic SOC; spin-orbit interaction opens a spectral gap at the Dirac points. In the case of graphene, the intrinsic gap equals~\cite{Gmitra2009:PRB} $2\lI\simeq 24\,\mu$eV. The spectral effects of the intrinsic SOC that are imprinted on the band structure are shown at Fig.~\ref{fig:d6h_li12}.

\subsection{No-go SOC matrix elements---lethal symmetries}\label{sec:no-go arguments}

In what follows we shortly summarize \emph{no-go arguments} showing explicitly how certain SOC mediated matrix elements become inhibited by specific structural symmetries. This will on one hand prove why for pristine graphene only the spin-conserving next nearest neighbor coupling $\lI$ is allowed. On the other hand, by seeing the absence of a particular \emph{no-go symmetry} in the symmetry group of a reduced hexagonal structure we can infer which additional coupling is allowed in the corresponding effective SOC Hamiltonian. We will profit from this insight in the forthcoming sections.
\subsubsection{Inhibition of all spin-flip SOCs---horizontal reflection}\label{subsubsec:inhibition by horizontal reflection}
Applying horizontal reflection $\sxy$ to a general spin-flip matrix element $\langle X_m\,\s|\Hh|X_n(-\s)\rangle$ between two $\pi$-states localized on arbitrary lattice sites $m$ and $n$, we get in accordance with Eq.~(\ref{eq:r_reflection_h}),
\begin{align}
\blangle &X_m\,\s\?\Hh\?X_n(-\s)\brangle=\nonumber\\
&= \blangle i(-1)^{\frac{1-\s}{2}}\sxy[X_m\,\s]\?\Hh\?i(-1)^{\frac{1+\s}{2}}\sxy[X_n(-\s)]\brangle\nonumber\\
&= -\blangle \sxy[X_m\,\s]\?\Hh\?\sxy[X_n(-\s)]\brangle\nonumber\\
&\overset{(\ref{Eq.:Unitary-Action})}{=} -\blangle X_m\,\s\?\Hh\?X_n(-\s)\brangle\,,\label{Eq.:inhibition-by-horizontal-reflection}
\end{align}
what implies that $\langle X_m\,\s|\Hh|X_n(-\s)\rangle=0$. Hence, we showed that the presence of $\sxy$ in the reduced point group inhibits any spin-flip terms in the effective SOC Hamiltonian. If $\sxy$ would not be present, then we would have a weaker result as discussed below.
\subsubsection{Inhibition of the nearest neighbor spin-flip SOCs---space-inversion, lattice translation, and time-reversal}\label{subsubsec:inhibition by inversion, translation and time-reversal}
For concreteness, let us focus on the SOC matrix element $\langle A_2\,\s\?\Hh\?B_3(-\s) \rangle$; see Fig.~\ref{fig:hex_sitelabel_coordinate}. Employing consecutively space-inversion $\mathcal{I}$, Eq.~(\ref{eq:r_spaceinversion}), unitarity, Eq.~(\ref{Eq.:Unitary-Action}), and translation by the lattice vector $\vec{a}=\overrightarrow{A_3A_2}=\overrightarrow{B_2B_3}$, Eq.~(\ref{eq:r_translation}), we get,
\begin{subequations}
\begin{align}
\blangle A_2\,\s\?\Hh\?B_3(-\s) \brangle &= \blangle -\mathcal{I}[B_2\,\s]\?\Hh\? -\mathcal{I}[A_3(-\s)] \brangle\nonumber\\
&= \blangle \mathcal{I}[B_2\,\s]\?\Hh\? \mathcal{I}[A_3(-\s)] \brangle\nonumber\\
&\overset{(\ref{Eq.:Unitary-Action})}{=} \blangle B_2\,\s\?\Hh\?A_3(-\s) \brangle\nonumber\\
&=\blangle T_{\vec{a}}[B_3\,\s]\?\Hh\? T_{\vec{a}}[A_2(-\s)] \brangle\nonumber\\
&\overset{(\ref{Eq.:Unitary-Action})}{=}\blangle B_3\,\s\?\Hh\? A_2(-\s) \brangle\,.\label{Eq.:inhibition-by-space-inversion-1}
\end{align}
To proceed further, we use the time-reversal symmetry, Eq.~(\ref{Eq.:opposite spin SOC elements}),
\begin{align}
\blangle B_3&\,\s\?\Hh\? A_2(-\s) \brangle&\overset{(\ref{Eq.:opposite spin SOC elements})}{=}-\blangle A_2\,\s\?\Hh\? B_3(-\s) \brangle\,.\label{Eq.:inhibition-by-space-inversion-2}
\end{align}
\end{subequations}
Combining Eqs.~(\ref{Eq.:inhibition-by-space-inversion-1}) and (\ref{Eq.:inhibition-by-space-inversion-2}) we immediately see that the nearest neighbor
SOC mediated spin-flip hopping $\langle A_2\,\s\?\Hh\?B_3(-\s) \rangle=0$. Repeating the same for the remaining neighboring lattice sites at Fig.~\ref{fig:hex_sitelabel_coordinate} we inhibit---by the space-inversion $\mathcal{I}$, lattice translation $T_{\vec{a}}$ and time-reversal $\mathcal{T}$---all other nearest neighbor spin-flip terms in the effective SOC Hamiltonian.
\subsubsection{Inhibition of the nearest neighbor spin-conserving SOCs---vertical reflection, and lattice translation}\label{subsubsec:inhibition by vertical reflection and translation}
By similar reasoning as above we can show that the SOC matrix element $\langle A_2\,\s\?\Hh\?B_3\,\s \rangle$ is zero whenever
lattice translation $T_{\vec{a}}$ and vertical reflection $\syz$ are present. Translation by the lattice vector $\vec{a}=\overrightarrow{A_2A_3}=\overrightarrow{B_3B_2}$ implies,
\begin{subequations}\label{Eq.:inhibition-by-vertical-reflection}
\begin{align}
\blangle A_2\,\s\?\Hh\?B_3\,\s \brangle &= \blangle T_{\vec{a}}[A_3\,\s]\?\Hh\? T_{\vec{a}}[B_2\,\s] \brangle\nonumber\\
&\overset{(\ref{Eq.:Unitary-Action})}{=} \blangle A_3\,\s\?\Hh\?B_2\,\s \brangle\,.
\end{align}
Moreover, using the vertical reflection, Eq.~(\ref{eq:r_reflection_yz}), we have $|A_3\,\s\rangle=-i\syz |A_2(-\s)\rangle$ and $|B_2\,\s\rangle=-i\syz |B_3(-\s)\rangle$ and therefore, by unitarity, Eq.~(\ref{Eq.:Unitary-Action}), we arrive at,
\begin{align}
\blangle A_3\,\s\?\Hh\?B_2\,\s \brangle &= \blangle A_2(-\s)\?\Hh\?B_3(-\s) \brangle\,.
\end{align}
So the last two equations together with Eq.~(\ref{Eq.:pure_imaginary2}) imply
\begin{align}
\blangle A_2\,\s\?\Hh\?B_3\,\s \brangle &= \blangle A_2(-\s)\?\Hh\?B_3(-\s) \brangle\nonumber\\
&\overset{(\ref{Eq.:pure_imaginary2})}{=} -\blangle A_2\,\s\?\Hh\?B_3\,\s \brangle\,,
\end{align}
\end{subequations}
which means that the nearest neighbor spin-conserving hopping $\langle A_2\,\s\?\Hh\?B_3\,\s \rangle$ is zero.
Repeating the same argumentation for the other neighboring sites we eliminate---by lattice translation $T_{\vec{a}}$
and vertical reflection $\syz$---all remaining nearest neighbor spin-conserving SOC terms.
\\

The \emph{no-go arguments} based on the horizontal and vertical reflections $\sxy, \syz\in D_{6h}$, see Eqs.~(\ref{Eq.:inhibition-by-horizontal-reflection}) and (\ref{Eq.:inhibition-by-vertical-reflection}), explain straightforwardly why the translationally invariant SOC Hamiltonian $\mathcal{H}_{D_{6h}}$ of pristine graphene, Eq.~(\ref{Eq.:pristine-graphene-SOC-Hamiltonian}), allows only the next nearest neighbor spin-conserving hoppings.

Pristine graphene is an example of a hexagonal system with the highest structural group symmetry.
The topic for the next sections are hexagonal systems with lower symmetries---subgroups of the point group $D_{6h}$. We will start with the maximal structural subgroups $D_{3d}$, $D_{3h}$, and $C_{6v}$---and explore step-by-step the symmetry allowed spin-orbit couplings.

\subsection{Subgroups of $D_{6h}$---categorization of emergent SOCs}\label{sec:subgroups of D6h}

\begin{figure*}
\includegraphics[width=1.5\columnwidth]{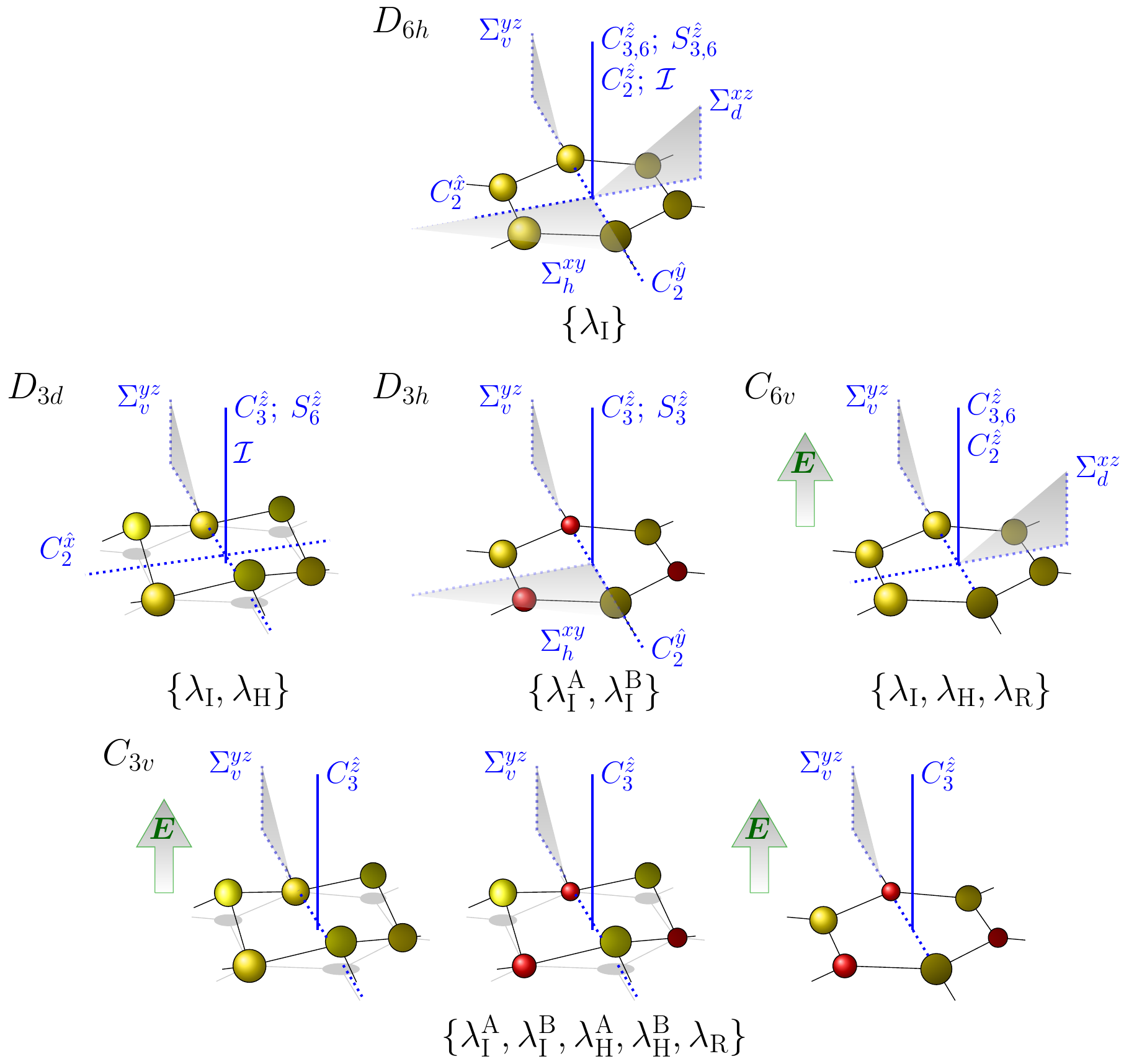}
\caption{(Color online) Point group $D_{6h}$ of pristine graphene and its maximal subgroups $D_{3d}$, $D_{3h}$ and $C_{6v}$. These are represented, for instance, by graphene mini-ripple, planar boron-nitride and graphene exposed to a transverse external electric field, respectively. The point groups $D_{3d}$, $D_{3h}$ and $C_{6v}$ share the common subgroup $C_{3v}$, as depicted (left to right) by the sample configurations of graphene mini-ripple in a transverse external electric field, mini-rippled boron-nitride and boron-nitride in a transverse external electric field. The successive reduction of the point group symmetry (top to bottom) enhances the number of symmetry-allowed SOC parameters.\label{fig:d6h_subgroups}}
\end{figure*}

\setlength{\tabcolsep}{6pt}
\begin{table*}
\newcolumntype{C}{>{\centering\arraybackslash}p{2.4em}}
\newcolumntype{D}{>{\centering\arraybackslash}p{4.3em}}
\begin{tabular}{cCCCCCCCCCCDDD}
\hline
Group$\bigl/$Operation     & $E$   & $2C_6^{\hat{z}}$ & $2C_3^{\hat{z}}$ & $C_2^{\hat{z}}$ & $3C_2'$ & $3C_2''$ & $\mathcal{I}$ & $\Sp_h$ & $3\Sp_v$ & $3\Sp_d$ & $2S_3^{\hat{z}}$ & $2S_6^{\hat{z}^{\phantom{A}}}$ \\
\hline\hline
 $D_{6h}^{\phantom{A}}$ & $\ch$ & $\ch$  & $\ch$  & $\ch$ &   $\ch$ & $\ch$    & $\ch$         & $\ch$   & $\ch$    & $\ch$    & $\ch$ & $\ch$ \\
 $D_{3d}$ & $\ch$ & ---    & $\ch$  & ---   &   $\ch$ & ---      & $\ch$         & ---     & $\ch$    & ---      & ---   & $\ch$ \\
 $D_{3h}$ & $\ch$ & ---    & $\ch$  & ---   &   ---   & $\ch$    & ---           & $\ch$   & $\ch$    & ---      & $\ch$ & ---   \\
 $C_{6v}$ & $\ch$ & $\ch$  & $\ch$  & $\ch$ &   ---   & ---      & ---           & ---     & $\ch$    & $\ch$    & ---   & ---   \\
 $C_{3v}$ & $\ch$ & ---    & $\ch$  & ---   &   ---   & ---      & ---           & ---     & $\ch$    & ---      & ---   & ---   \\
\hline
\end{tabular}
\caption{Point group $D_{6h}$ and its maximal subgroups---$D_{3d}$, $D_{3h}$, $C_{6v}$---including also their common intersection---the point group $C_{3v}$. For the visualization see Fig.~\ref{fig:d6h_subgroups}. We shortened the notation in terms of the previous definitions: symbol $2C_6^{\hat{z}}$ means two 6-fold rotations along the $z$-axis, namely $\Rz_{\pm\pi/3}$, symbol $3C_2'$ stands for three 2-fold rotations along the axis
$x$, $\Rz_{\pi/3}x$ and $\Rz_{2\pi/3}x$, respectively, and similarly $3C_2''$ stands for three 2-fold rotations along the $y$, $\Rz_{\pi/3}y$ and $\Rz_{2\pi/3}y$ axis, respectively. By the same logic, $3\Sp_v$ stands for three mirror reflections in $yz$, $\Rz_{\pi/3}yz$ and $\Rz_{2\pi/3}yz$ planes, respectively, and so on. If the given set of operations is present/absent in the particular subgroup of $D_{6h}$ we employ the marker $\ch$/---.}\label{Tab:all-data}
\end{table*}

Any periodic modification of the pristine hexagonal symmetry reduces the unit cell point group symmetry $D_{6h}$ to one of its subgroups and is
manifested by the emergence of new SOC mediated hoppings. The aim of this section is to show a bottom line enabling their classification and categorization.

The minimal structural modifications we will discuss here are (1)\,\emph{rippling}, (2)\,\emph{sublattice asymmetry}, and (3)\,\emph{transverse electric field or substrate} and their mutual combinations, see Fig.~\ref{fig:d6h_subgroups}.
We call here a structural modification of the full hexagonal lattice \emph{minimal}, if the reduced point subgroup of $D_{6h}$ experiences minimal modifications
in terms of the number of group elements. Such subgroups are usually called \emph{maximal subgroups}. In the case of $D_{6h}$ there are in total five maximal subgroups.\cite{KDWS1963:MIT} Each of them has 12 group elements---\emph{group order} 12---which is half of the order of the original point group $D_{6h}$. Three subgroups---$D_{3d}$, $D_{3h}$, $C_{6v}$---will be relevant in the present context, while the subgroups $D_6$ and $C_{6h}$ are irrelevant for us. To be specific:
\begin{itemize}
\item \emph{rippling} reduces $D_{6h}\rightarrow D_{3d}$ what constitutes the point group of graphene mini-ripple, graphane\cite{Sofo2007:PRB}, silicene and `gelicene'\cite{Takeda1994:PRB,Cahangirov2009:PRL}, etc.;
\item \emph{sublattice inversion asymmetry} reduces $D_{6h}\rightarrow D_{3h}$ what is the point group of the planar boron-nitride, aluminum-nitride, or any other planar system with two non-equivalent interpenetrating triangular lattices $A$ and $B$;
\item \emph{transverse electric field} reduces $D_{6h}\rightarrow C_{6v}$ what represents the point group of pristine graphene in an external field or graphene deposited on a substrate that is not breaking the sublattice symmetry.
\end{itemize}
For visualization, summary, and mutual comparison see Fig.~\ref{fig:d6h_subgroups} and Table~\ref{Tab:all-data}.

It is worth to emphasize that an intersection of any two of $D_{3d}$, $D_{3h}$, $C_{6v}$ is isomorphic\cite{KDWS1963:MIT} to the smaller non-abelian subgroup $C_{3v}\subset D_{6h}$ with group order $6$.
This means that an arbitrary combination of two minimal structural modifications leads to the same effective SOC Hamiltonian, which possesses global $C_{3v}$ invariance. For concreteness, graphene mini-ripple (or graphane, silicene, gelicene) in a transverse electric field---$D_{3d}\cap C_{6v}$---is from the effective SOC point of view equivalent to a mini-rippled boron-nitride without the field or free standing graphone\cite{Zhou2009:NanoLett}---$D_{3d}\cap D_{3h}$.

With respect to the structural minimality the point groups $D_{3d}$, $D_{3h}$, and $C_{6v}$ can be considered as
equivalent since they are all maximal subgroups of $D_{6h}$. Despite of that minimal subgroup similarity, $D_{3d}$, $D_{3h}$, and $C_{6v}$ are different since they result in different SOC phenomena.
\subsubsection{$D_{3d}$-case: $\lI$ and $\lH$ couplings}
Rippled structures such as graphane, silicene, and graphene mini-ripple---point group $D_{3d}$---remain invariant under the space-inversion $\mathcal{I}$ and time-reversal $\mathcal{T}$, and hence SOC can not cause band spin splittings.
The reasoning is finger counting:\cite{Fabian2007:APS} for any band index $n$ we have
\begin{equation}\label{eq:no-splitting}
\varepsilon_{n,\s}(\mathbf{k})\overset{\mathcal{T}}{=}\varepsilon_{n,-\s}(-\mathbf{k})\overset{\mathcal{I}}{=}\varepsilon_{n,-\s}(\mathbf{k})\,.
\end{equation}
Space-inversion $\mathcal{I}$ and vertical reflection $\syz$ belong to $D_{3d}$, but the horizontal reflection $\sxy$ does not. Then, according to the \emph{no-go arguments} presented in section~\ref{sec:no-go arguments}, the $D_{3d}$ symmetric and time-reversal invariant SOC Hamiltonian based on $\pi$-orbitals allows only next nearest neighbor hoppings. The nearest neighbor SOCs are inhibited---the spin-conserving ones by $\syz$ and spin-flipping by $\mathcal{I}$. Because $\mathcal{I}$ interchanges sublattices, $|A_i\,\s\rangle=-\mathcal{I}|B_i\,\s\rangle$, see Fig.~\ref{fig:hex_sitelabel_coordinate}
and Eq.~(\ref{eq:r_spaceinversion}), the next nearest hoppings should not be sublattice resolved. Indeed,
\begin{align}
\blangle A_i\,\s\?\Hh\?A_j\,\s' \brangle &\overset{(\ref{Eq.:Unitary-Action})}{=} \blangle B_i\,\s\?\Hh\?B_j\,\s' \brangle\,.
\end{align}
Similarly, $\mathcal{T}$ interchanges the spin components,
\begin{align}
\blangle X_m\,&\s\?\Hh\?X_n\,\s' \brangle\nonumber\\
&\overset{(\ref{Eq.:opposite spin SOC elements})}{=}-(-1)^{\frac{\s+\s'}{2}}\, \blangle X_n(-\s')\?\Hh\?X_m(-\s) \brangle\,,
\end{align}
and hence there is only one purely imaginary spin-conserving hopping, say, defined for $\s'=\s=\ \ua$, and one spin-flipping hopping defined for $\s'=-\s=\ \da$, respectively.

It is now a convention---by analogy with the plain graphene---to call the spin-conserving next nearest neighbor SOC matrix element intrinsic. Hence also in the $D_{3d}$ case we adopt the term intrinsic SOC. We define intrinsic $i\lI$ by the same prescription as already given by Eq.~(\ref{Eq.:intr_graphene}):
\begin{align}\label{Eq.:intr_D3d}
\frac{i\lI}{3\sqrt{3}}&=\blangle A_3\ua\?\Hh\?A_2\ua \brangle=\blangle B_2\da\?\Hh\?B_3\da \brangle\,.
\end{align}
The related sublattice-spin sign factors are governed by the prefactor $\nu_{m,n}^{\phantom\dagger}\,[\hat{s}_z]_{\sigma\sigma}$ as discussed above.

There is no terminological consensus on how to call the spin-flipping next nearest neighbor SOC matrix element. Such a term already emerged in bilayer graphene\cite{Konschuh2012:PRB}, but that time its group symmetry origin was not discussed. Later, when studying SOC effects in semi-hydrogenated graphene (graphone) the acronym PIA---a shorthand for the ``pseudospin inversion asymmetry'' was proposed\cite{Gmitra2013:PRL}. In that case, the pseudospin was explicitly broken by the hydrogenation of one sublattice resulting in the $C_{3v}$ invariant structure. Unfortunately, the pseudospin asymmetry is not supported by the point group $D_{3d}$ which contains the space-inversion $\mathcal{I}$. So the former PIA acronym is not fully appropriate in $D_{3d}$ case. Alternatively, authors of Ref.~[\onlinecite{Liu2011:PRB}] used the term ``intrinsic Rashba SOC''. This is also inappropriate, since normally the Rashba\cite{Bychkov-Rashba1983:JETP-Letters} SOC causes band splittings and this is also not the case in $D_{3d}$ invariant systems.

The emergence of the spin-flipping next nearest neighbor SOC is related to the absence of the horizontal reflection $\sxy$ in the underlying point group (see also other cases discussed below). Since the horizontal plane is a principal mirror plane of the structure we can call it ``Principal-plane mIrror Asymmetry'' induced SOC, preserving the subscript PIA (by explicitly breaking with abbreviation rule). Thus the PIA spin-orbit coupling $\lH$ can be defined as:
\begin{align}\label{Eq.:PIA_D3d}
\frac{2}{3}\lH&\equiv\blangle A_3\ua\?\Hh\?A_2\da \brangle\overset{(\ref{Eq.:opposite spin SOC elements})}{=}-\blangle A_2\ua\?\Hh\?A_3\da \brangle\,.
\end{align}
Again, the numerical prefactor $2/3$ is a matter of convenience.
Employing the vertical reflection $\syz\in D_{3d}$ we show that $\lH$ is purely real
\begin{align}\label{Eq.:D3d pure real lH}
\blangle A_3\ua&\?\Hh\?A_2\da \brangle\overset{(\ref{eq:r_reflection_yz})}{=}\blangle -i\syz[A_2\da]\?\Hh\? -i\syz[A_3\ua] \brangle\nonumber\\
&\overset{(\ref{Eq.:Unitary-Action})}{=}\blangle A_2\da\?\Hh\? A_3\ua \brangle
=\overline{ \blangle A_3\ua\?\Hh\?A_2\da \brangle}\,.
\end{align}
As a consequence of the last two equations we have the practical identity:
\begin{align}
\frac{2}{3}\lH=\blangle A_3\ua\?\Hh\?A_2\da \brangle=-\blangle A_3\da\?\Hh\?A_2\ua \brangle\,.
\end{align}
The remaining next nearest neighbor spin flipping SOCs on the $A$-sublattice, see Fig.~(\ref{fig:hex_sitelabel_coordinate}), can be connected with $\lH$ by
rotations $\mathcal{R}^{\hat{z}}_{\pm\frac{2\pi}{3}}\in D_{3d}$. In particular we get,
\begin{align}
\blangle A_1\ua\?\Hh\?A_3\da \brangle
&\overset{(\ref{eq:r_rotation})}{=}e^{-i\frac{2\pi}{3}}\blangle \mathcal{R}^{\hat{z}}_{\frac{2\pi}{3}}[A_3\ua]\?\Hh\? \mathcal{R}^{\hat{z}}_{\frac{2\pi}{3}}[A_2\da] \brangle\nonumber\\
&\overset{(\ref{Eq.:Unitary-Action})}{=}e^{-i\frac{2\pi}{3}}\,\frac{2}{3}\lH\,,\\
\blangle A_2\ua\?\Hh\?A_1\da \brangle
&\overset{(\ref{eq:r_rotation})}{=}e^{i\frac{2\pi}{3}}\blangle \mathcal{R}^{\hat{z}}_{-\frac{2\pi}{3}}[A_3\ua]\?\Hh\? \mathcal{R}^{\hat{z}}_{-\frac{2\pi}{3}}[A_2\da] \brangle\nonumber\\
&\overset{(\ref{Eq.:Unitary-Action})}{=}e^{i\frac{2\pi}{3}}\,\frac{2}{3}\lH\,.
\end{align}
The SOC matrix elements on the sublattice $B$ can be obtained from the above $A$-sublattice formulas after employing the space-inversion.
The spin-flipping next nearest neighbor SOC elements for both sublattices can be compactly summarized by the following formula,
\begin{equation}\label{Eq.:D3d phase factors}
\blangle X_m\,\s\,\?\Hh\? X_n\,\s'\brangle
=\bigl[i\hat{\boldsymbol{s}}\times\boldsymbol{\mathrm{d}}_{m,n}\bigr]_{\s\s'}\,\frac{2}{3}\,\lH\,,
\end{equation}
where $\boldsymbol{\mathrm{d}}_{m,n}=\overrightarrow{mn\ }\bigl/|\overrightarrow{mn\ }|$ is the unit vector in the horizontal ($xy$) plane pointing from the lattice site $n$ to the next nearest neighbor site $m$; $\hat{\boldsymbol{s}}$ stands for the array of Pauli matrices and spin projections $\s\neq\s'$.

\begin{figure}
\includegraphics[width=1.0\columnwidth]{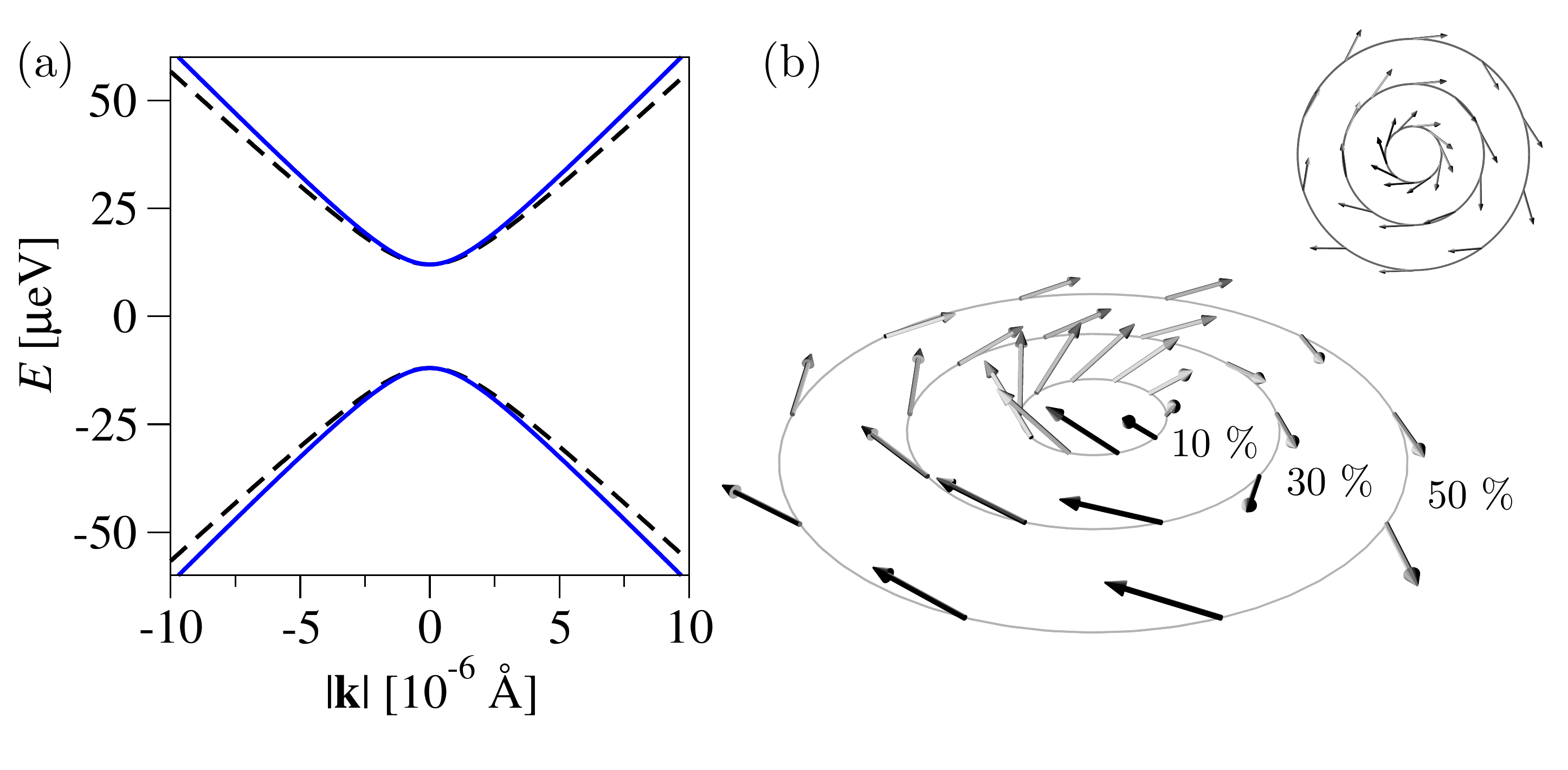}
\caption{(Color online) Electronic band structure and spin-orbit field around the K-point in the presence of $D_{3d}$ invariant SOC Hamiltonian.
Panel (a): electronic band structure for $t = 2.6\,\mathrm{eV}$, $\lI = 12 \,\mu\mathrm{eV}$ and $\lH = 0.1\,\mathrm{eV}$ (black dashed) and $\lH = 1\,\mathrm{eV}$ (blue solid), respectively, showing the effect of renormalization of Fermi velocity. Panel (b): $D_{3d}$ spin-orbit field around the Dirac
point along the circles with radius 10\%, 30\%, and 50\% of $\overline{\mathrm{KM}}$-distance for $t = 2.6\,\mathrm{eV}$, $\lI = 12 \,\mu\mathrm{eV}$ and
$\lH = 60\,\mu\mathrm{eV}$. The inset shows a top view of the spin-orbit field. The circular low energy symmetry changes to the triangular one and the
$z$-component of the spin-orbit field becomes suppressed when moving away from the Dirac point.
\label{fig:d3d_field}}
\end{figure}

To summarize, the effective translationally invariant SOC Hamiltonian based on $\pi$-orbitals that respects $D_{3d}$ symmetry and time-reversal is given by,
\begin{align}\label{Eq.:D3d-SOC-Hamiltonian}
\mathcal{H}_{D_{3d}}&=
\frac{i\lI}{3\sqrt{3}}\sum\limits_\sigma\sum\limits_{\llangle m,n\rrangle}
\nu_{m,n}^{\phantom\dagger}\,\bigl[\hat{s}_z\bigr]_{\sigma\sigma}\,\?X_m\,\s\brangle\,\blangle X_n\,\s\?\\
&+\frac{2\lH}{3}\sum\limits_{\s\neq\s'}\sum\limits_{\llangle m,n\rrangle} \bigl[i\hat{\boldsymbol{s}}\times\boldsymbol{\mathrm{d}}_{m,n}\bigr]_{\s\s'}\,\?X_m\,\s\brangle\,\blangle X_n\,\s'\?\,.\nonumber
\end{align}
Transforming the above SOC Hamiltonian into the Bloch form, $\mathcal{H}_{D_{3d}}=\sum_{\mathbf{q}}\mathcal{H}_{D_{3d}}(\mathbf{q})$, we arrive at,
\begin{align}\label{Eq.:D3d-SOC-Hamiltonian-Bloch Form}
&\mathcal{H}_{D_{3d}}(\mathbf{q})=\sum\limits_{X,\s,\s'}\bigl[\hat{\s}_z\bigr]_{XX}\,
\Bigl\{
\lI f_{\mathrm{I}}(\mathbf{q})\bigl[\hat{s}_z\bigr]_{\s\s'}\,+\\
&\lH f_{\mathrm{P}}(\mathbf{q})\bigl[\hat{s}_{+}\bigr]_{\s\s'}+\lH \overline{f_{\mathrm{P}}(\mathbf{q})}\bigl[\hat{s}_{-}\bigr]_{\s\s'}\Bigr\}\,\?X_{\mathbf{q}}\,\s\brangle\blangle X_{\mathbf{q}}\,\s'\?
\nonumber\,,
\end{align}
where the structural SOC function $f_{\mathrm{I}}(\mathbf{q})$ is given by Eq.~(\ref{Eq:structural-I-funcion}) and $f_{\mathrm{P}}(\mathbf{q})$ is defined as follows
\begin{align}\label{Eq:structural-H-funcion}
f_{\mathrm{P}}(\mathbf{q})=\frac{4i}{3}\Bigl\{&\sin{\mathbf{q}\cdot\mathbf{R}_1}\,+&\\
&\ +\mathrm{e}^{-i\frac{2\pi}{3}}\sin{\mathbf{q}\cdot\mathbf{R}_2}+\mathrm{e}^{+i\frac{2\pi}{3}}\sin{\mathbf{q}\cdot\mathbf{R}_3}\Bigr\}\,.\nonumber
\end{align}
A direct inspection shows that the quasi-momentum dependent spin operator [in units of $\hbar/2$]
\begin{align}
\mathrm{Spin}(\mathbf{q})&=\hat{\s}_0\,\Bigl[f_{\mathrm{P}}(\mathbf{q})\,\hat{s}_{+}
+\overline{f_{\mathrm{P}}(\mathbf{q})}\,\hat{s}_{-}
+\frac{\lI}{\lH}\,f_{\mathrm{I}}(\mathbf{q})\,\hat{s}_z\Bigr]
\end{align}
commutes with $\mathcal{H}_{D_{3d}}(\mathbf{q})$. Since the orbital Hamiltonian is diagonal in spin space, the eigenstates of
$H_{\rm orb}(\mathbf{q})+\mathcal{H}_{D_{3d}}(\mathbf{q})$ can be chosen as ``spin-up'' and ''spin-down'' states with respect to the momentum
dependent quantization axis specified by the unit vector:
\begin{align}\label{Eq:D3d SOC field}
\mathbf{n}(\mathbf{q})&=
\frac{
\left(\mathrm{Re}\bigl[f_{\mathrm{P}}(\mathbf{q})\bigr],-\mathrm{Im}\bigl[f_{\mathrm{P}}(\mathbf{q})\bigr],\frac{\lI}{\lH}f_{\mathrm{I}}(\mathbf{q})
\right)}
{\sqrt{\bigl|f_{\mathrm{P}}(\mathbf{q})\bigr|^2+\frac{\lI^2}{\lH^2}\,\bigl|f_{\mathrm{I}}(\mathbf{q})\bigr|^2}}\,.
\end{align}
Consequently, $\mathrm{Spin}(\mathbf{q})\simeq\hat{\s}_0\bigl[\mathbf{n}(\mathbf{q})\cdot{\hat{\boldsymbol{s}}}\bigr]$.
It is clear that at the time-invariant momenta, i.e.,~at $\Gamma$ and $\rm M$ points, $\mathbf{n}(\mathbf{q})$ is not well-defined. Hence there is not a well-defined global map from the full first Brillouin zone ($2d$ torus) into the $2d$ sphere, $\mathbf{q}\mapsto\mathbf{n}(\mathbf{q})$, and thus not a well-defined global winding number. Expanding $f_{\mathrm{P}}(\mathbf{q})$ around the Dirac points, $\mathbf{q}=\tau\mathbf{K}+\mathbf{k}$, keeping the first appearing non-zero terms we get,
\begin{equation}
f_{\mathrm{P}}(\tau\mathbf{K}+\mathbf{k})\simeq -(i k_x+k_y) a_{\mathrm{L}}\,,
\end{equation}
where $a_{\mathrm{L}}$ stands for the lattice constant. Then the effective $D_{3d}$-invariant low energy Hamiltonian around $\tau\mathbf{K}$-valley that includes both orbital and SOC terms is given by,
\begin{align}\label{Eq.:Effective-D3d}
\mathcal{H}_{\mathrm{eff}}(\tau\mathbf{K}+\mathbf{k})&=\hbar v_F\,\bigl(\tau k_x\hat{\s}_x-k_y\hat{\s}_y\bigr)\hat{s}_0+\tau\lI\hat{\s}_z\hat{s}_z+\nonumber\\
&+\lH\hat{\s}_z(k_x \hat{s}_y - k_y \hat{s}_x)a_{\mathrm{L}}\,.
\end{align}
Correspondingly, the momentum dependent spin quantization axis is aligned along the unit vector,
\begin{equation}
\mathbf{n}(\tau\mathbf{K}+\mathbf{k})=\frac{\bigl(-k_y a_{\mathrm{L}}, k_x a_{\mathrm{L}},\tau\tfrac{\lI}{\lH}\bigr)}{\sqrt{(k_x^2+k_y^2)a_{\mathrm{L}}^2+\frac{\lI^2}{\lH^2}}}\,.
\end{equation}
The eigenspectrum of $\mathcal{H}_{\mathrm{eff}}(\tau\mathbf{K}+\mathbf{k})$---labeled by quasi-momentum $\mathbf{k}$, band index $n=\pm$ and spin $\sigma$ with respect to $\mathbf{n}(\tau\mathbf{K}+\mathbf{k})$---reads:
\begin{equation}
\varepsilon_{n,\s}(\tau\mathbf{K}+\mathbf{k})=n\sqrt{\lI^2+\left(\hbar^2 v_F^2+\lH^2 a_{\mathrm{L}}^2\right) (k_x^2+k_y^2)}\,,
\end{equation}
so the states are indeed spin degenerate as we already noticed in Eq.~(\ref{eq:no-splitting}). The effect of $\lH$ SOC is twofold. First, looking at the eigenspectrum, $\lH$ effectively renormalizes the Fermi velocity $v_F\rightarrow \sqrt{v_F^2+\lH^2 a_{\mathrm{L}}^2/\hbar^2}$, or equivalently in terms of the orbital nearest neighbor hopping $t\rightarrow \sqrt{t^2+4\lH^2/3}$.
In situations when the strength of orbital hopping $t$ substantially exceeds the strength of $\lH$ this effect is expected to be marginal,
e.g.~in silicene\cite{Liu2011:PRB} $t\simeq 1.1$\,eV and $\lH\simeq 0.7$\,meV, however in gelicene\cite{Liu2011:PRB}
$t\simeq 0.9$\,eV and $\lH\simeq 10.7$\,meV and hence the renormalization of the orbital hopping should be more pronounced. Second, $\lH$ introduces a non-trivial spin-orbit field in the k-space, $\mathbf{n}(\mathbf{q})\cdot{\hat{\boldsymbol{s}}}$, that gives rise to the in-plane component of the spin-expectation value, see
Fig.~\ref{fig:d3d_field}.
\subsubsection{$D_{3h}$-case: $\lIA$ and $\lIB$ couplings}
\begin{figure}
\includegraphics[width=1.\columnwidth]{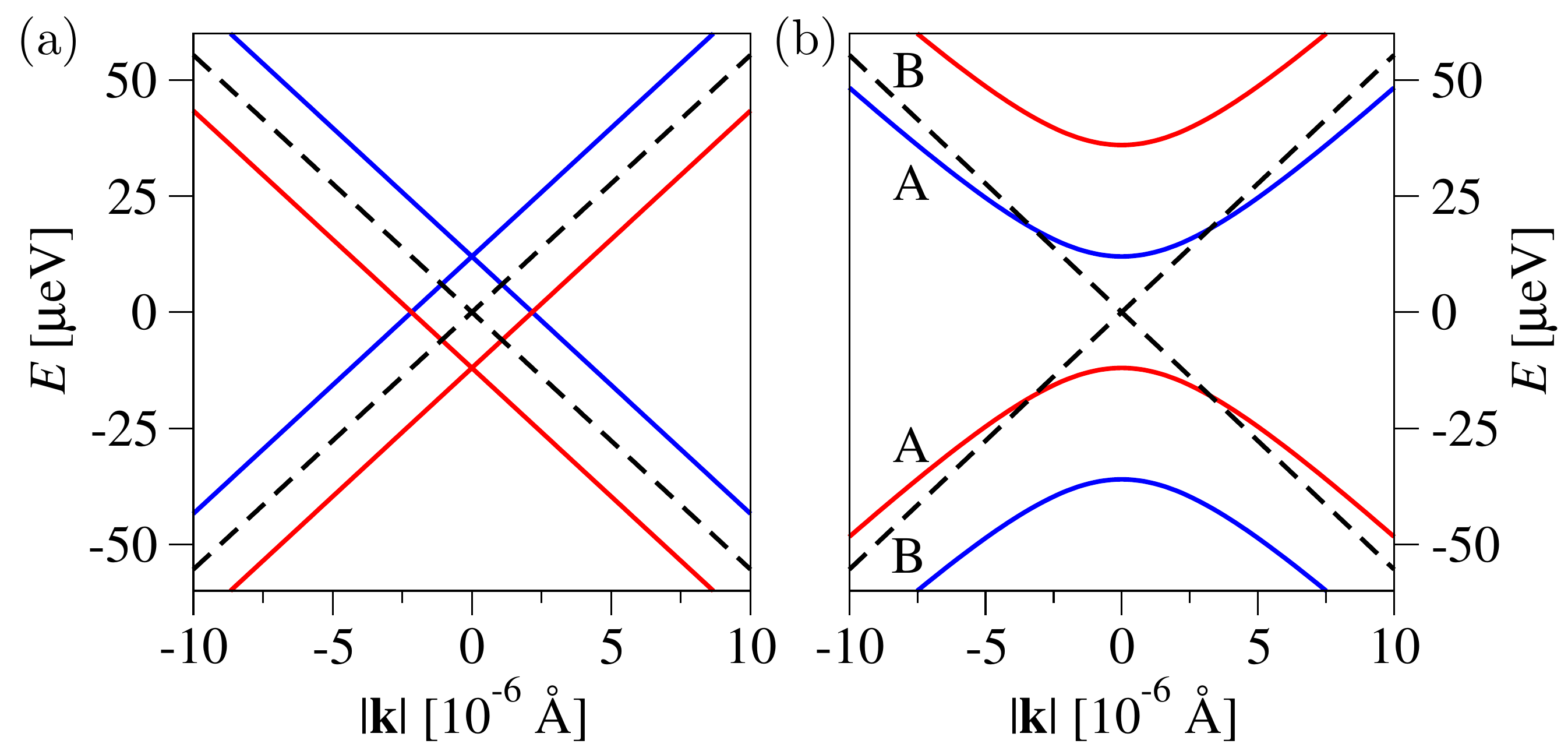}
\caption{(Color online)
Electronic band structure around the K-point in the presence of $D_{3h}$ invariant SOC Hamiltonian without staggered potential $\Delta$; for
$t=2.6$\,eV and $\lIA= 12\,\mu$eV and (a) $\lIB=-\lIA$ and (b) $\lIB=3\lIA$. Blue and red lines indicate bands with up and down spin projections, respectively.
Label A or B at the given band indicates which sublattice is dominantly occupied by electronic states at that band, assuming their momenta are close to the Dirac point. For comparison, the black dashed lines display the energy dispersion of the pristine graphene without SOC and the staggered $\Delta$.
\label{fig:d3h_lia12mlib_lb3lia}}
\end{figure}

Hexagonal boron-nitride is a prototype of a planar hexagonal structure that consists of two non-equivalent interpenetrating triangular lattices---in our particular example composed of borons and nitrogens, respectively. Since the horizontal reflection $\sxy$ belongs to $D_{3h}$, spin-flipping SOC mediated hoppings are not allowed according to the \emph{no-go argument}~\ref{subsubsec:inhibition by horizontal reflection}.
Similarly, the vertical mirror reflection $\syz$ belongs to $D_{3h}$ and hence by the assertion \ref{subsubsec:inhibition by vertical reflection and translation} there are neither spin-conserving nearest neighbor SOCs. Therefore, we are left with the intrinsic---next nearest neighbor spin-conserving SOC---terms only. The broken sublattice symmetry can not further constrain the intrinsic SOCs and hence they become sublattice dependent, i.e.~$\lIA\neq\lIB$. Motivated by the previous analysis and knowing that they are purely imaginary we define them via the formulas
\begin{align}\label{Eq.:intr_D3h}
\frac{i\lIA}{3\sqrt{3}}&=\blangle A_3\ua\?\Hh\?A_2\ua \brangle, \,\\
\frac{i\lIB}{3\sqrt{3}}&=\blangle B_2\da\?\Hh\?B_3\da \brangle\,;
\end{align}
for the atomic sites configuration see Fig.~\ref{fig:hex_sitelabel_coordinate}. In analogy with Eq.~(\ref{Eq.:pristine-graphene-SOC-Hamiltonian}) the $D_{3h}$ invariant SOC Hamiltonian reads,
\begin{align}\label{Eq.:D3h-SOC-Hamiltonian}
\mathcal{H}_{D_{3h}}&=
\frac{i\lIA}{3\sqrt{3}}\sum\limits_\sigma\sum\limits_{\llangle m,n\rrangle}
\nu_{m,n}^{\phantom\dagger}\,\bigl[\hat{s}_z\bigr]_{\sigma\sigma}\,\?A_m\,\s\brangle\,\blangle A_n\,\s\?\nonumber\\
&+
\frac{i\lIB}{3\sqrt{3}}\sum\limits_\sigma\sum\limits_{\llangle m,n\rrangle}
\nu_{m,n}^{\phantom\dagger}\,\bigl[\hat{s}_z\bigr]_{\sigma\sigma}\,\?B_m\,\s\brangle\,\blangle B_n\,\s\?\,.
\end{align}
Contrary to the $D_{6h}$ case, the lack of space-inversion symmetry $\mathcal{I}$ in $D_{3h}$---hence two different values of $\lIA$ and $\lIB$---causes spin splitting of the band structure, see Fig.~\ref{fig:d3h_lia12mlib_lb3lia}.

\begin{figure}
\includegraphics[width=1.\columnwidth]{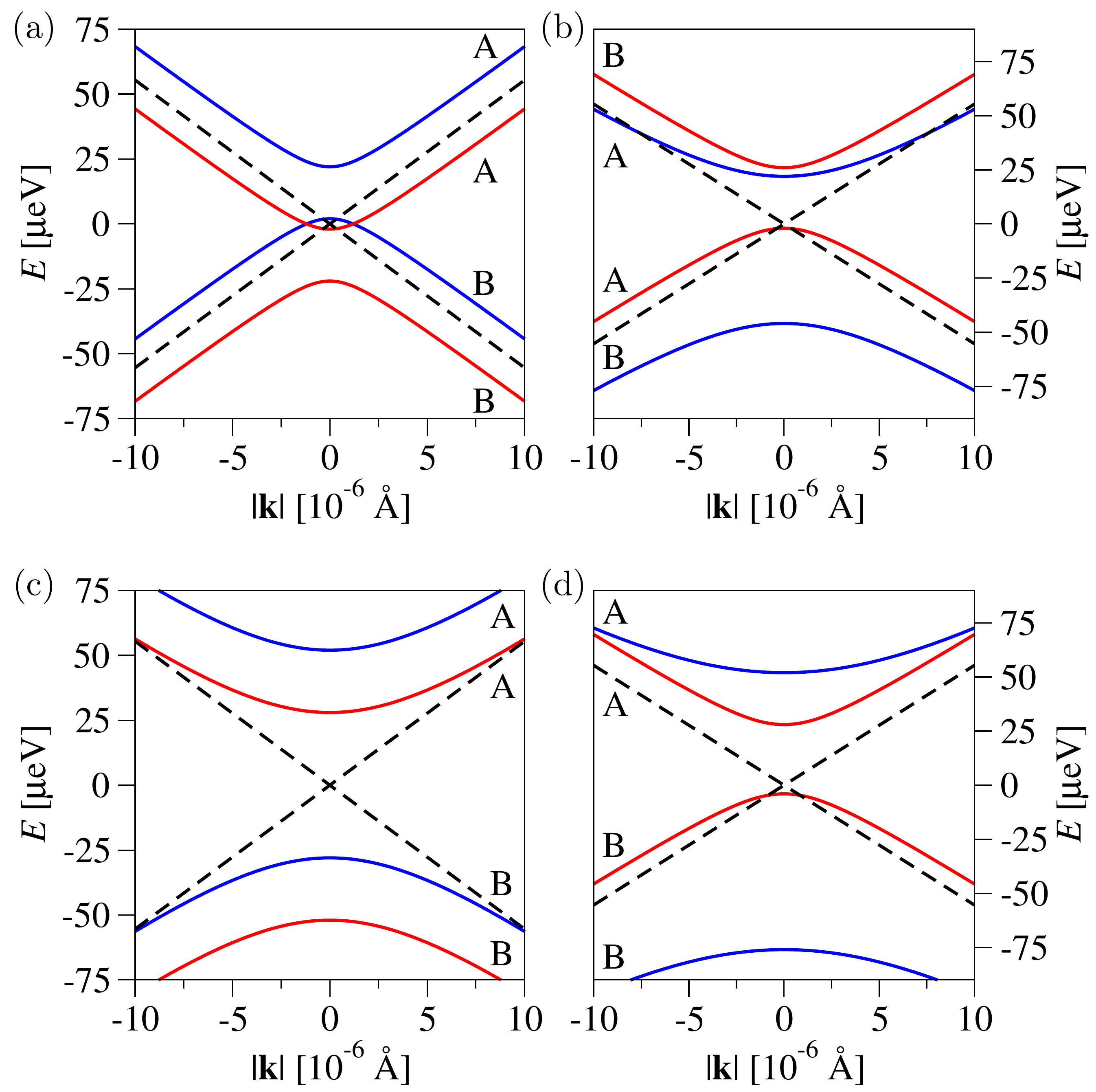}
\caption{(Color online)
Electronic band structure around the K-point in the presence of $D_{3h}$ invariant SOC Hamiltonian and staggered potential $\Delta$
for $t=2.6$\,eV and $\lIA= 12\,\mu$eV and (a) $\lIB=-\lIA$ and $\Delta = 10\,\mu$eV, (b) $\lIB=3\lIA$ and $\Delta = 10\,\mu$eV, (c) $\lIB=-\lIA$ and $\Delta = 40\,\mu$eV, and (d) $\lIB=3\lIA$ and $\Delta = 40\,\mu$eV, respectively.
For $\mathrm{sign}\,\lIA \neq \mathrm{sign}\,\lIB$ the increased value of the staggered potential drives the band-inverted structure, panel (a), to the insulating one, panel (c). Blue and red lines indicate bands with up and down spin projections, respectively, and the dashed lines display the orbital band structure of pristine graphene. Label A or B at the given band indicates which sublattice is dominantly occupied by electronic states at that band, assuming their momenta are close to the Dirac point.
\label{fig:d3h_lia12mlib_lb3lia:part2}}
\end{figure}

The low energy Bloch representation of $\mathcal{H}_{\mathrm{orb}}(\tau\mathbf{K}+\mathbf{k})+\mathcal{H}_{D_{3h}}(\tau\mathbf{K}+\mathbf{k})$ can be easily deduced from Eq.~(\ref{Eq.:Effective-D6h}) when properly substituting $\lI$ by its sublattice resolved counterparts $\lIA$ and $\lIB$. The result is as follows
\begin{align}\label{Eq.:Effective-D3h}
\mathcal{H}_{\mathrm{eff}}(\tau\mathbf{K}+\mathbf{k})&=\hbar v_F\,\bigl(\tau k_x\hat{\s}_x-k_y\hat{\s}_y\bigr)\hat{s}_0+\Delta\,\hat{\s}_z\hat{s}_0+\\
&+\tfrac{\tau}{2}\left[ \lIA(\hat{\s}_z+\hat{\s}_0) + \lIB(\hat{\s}_z-\hat{\s}_0)\right ]\hat{s}_z\,.\nonumber
\end{align}
Contrary to the previous cases, the broken sublattice symmetry allows also a new term in the orbital Hamiltonian $\mathcal{H}_{\mathrm{orb}}$---the second term in the first line parameterized by the so called staggered potential $\Delta$. The two inequivalent sublattices can possess different on-site energies and their difference equals $2\Delta$. Similarly as in the $D_{6h}$ case, the spin operator $\hat{\s}_0\,\hat{s}_z$ commutes with $\mathcal{H}_{\mathrm{eff}}$ allowing us to label its eigenstates with the spin up and spin down entries. The eigenspectrum of $\mathcal{H}_{\mathrm{eff}}(\tau\mathbf{K}+\mathbf{k})$---labeled by the quasi-momentum $\mathbf{k}$, conduction/valence band index $n=+\bigl/-$ and spin $\sigma=\{\ua,\da\}=\{+1,-1\}$ with respect to $\hat{s}_z$ reads:
\begin{align}\label{Eq.:D3h-eigenspectrum}
\varepsilon_{n,\s}(\tau \mathbf{K}&+\mathbf{k})=\tfrac{\s}{2}(\lIA-\lIB)\,+\\
&+n\sqrt{\left[\Delta+\tfrac{\s}{2}(\lIA+\lIB)\right]^2+\hbar^2v_F^2(k_x^2+k_y^2)}\,.\nonumber
\end{align}
The band structure visualization of the SOC induced splittings in the presence of staggered $\Delta$ are displayed in Fig.~\ref{fig:d3h_lia12mlib_lb3lia:part2}. Direct analysis of Eq.~(\ref{Eq.:D3h-eigenspectrum}) shows that there are two distinct spectral cases---an insulating (gapped)
and a band-inverted (gapless) one.
The \emph{criterium to get spectral band-inversion} is, $\mathrm{sign}\,\lIA \neq \mathrm{sign}\,\lIB$, and $|\Delta|<\max(|\lIA|,|\lIB|)$.

\subsubsection{$C_{6v}$-case: $\lI$, $\lR$, and $\lH$ couplings}
Graphene in an external transverse electric field or graphene disposed on a substrate is a prototype of the structure with $C_{6v}$ structural symmetry. In this case the sublattices remain equivalent---the rotation $\mathcal{R}^{\hat{z}}_{\pi/3}$ that interchanges them belongs to the point group. However, we lose all the structural symmetries flipping the orientation of the transverse $z$-axis. According to the arguments in section~\ref{sec:no-go arguments},
lack of space-inversion $\mathcal{I}$ and horizontal reflection $\sxy$ can not prevent the system from spin-flip SOC hoppings among nearest and next nearest neighbors.
Contrary to that, $\syz$ inhibits the spin-conserving nearest neighbor SOCs, but allows intrinsic---next nearest neighbor---SOCs. From this finger counting symmetry analysis and the \emph{no-go arguments} we know that the $C_{6v}$ invariant SOC Hamiltonian would potentially host three couplings: $\lI$ and $\lH$---the terms identical with the already discussed $D_{3d}$ case, see Eq.~(\ref{Eq.:intr_D3d}), (\ref{Eq.:PIA_D3d}) and (\ref{Eq.:D3d-SOC-Hamiltonian})---and the new spin-flipping term $\lR$ acting between the nearest neighbors. Conventionally the latter is called Rashba SOC\cite{Kane_Mele_1_2005:PRL} and in terms of a SOC matrix element it can be defined as follows,
\begin{align}\label{Eq.:BR C6v}
\frac{2}{3}i\lR&\equiv
\blangle A_2\ua\?\Hh\?B_3\da\brangle
\overset{(\ref{Eq.:opposite spin SOC elements})}{=}
-\blangle B_3\ua\?\Hh\?A_2\da\brangle\,.
\end{align}
In the above definition we have already employed the purely imaginary character of the coupling. Applying the dihedral reflection $\sxz$ to the defining matrix element we obtain,
\begin{align}\label{Eq.:pure_imaginary_Rashba}
\blangle A_2&\ua\?\Hh\?B_3\da\brangle=\blangle \sxz[B_3\da]\?\Hh\? -\sxz[A_2\ua]\brangle\nonumber\\
&\overset{(\ref{Eq.:Unitary-Action})}{=}-\blangle B_3\da\?\Hh\? A_2\ua\brangle=-\overline{\blangle A_2\ua\?\Hh\? B_3\da\brangle}\,,
\end{align}
what is indeed what we wanted to show. In analogy with Eq.~(\ref{Eq.:D3d phase factors}) we can also write a compact formula for any nearest neighbor spin-flipping matrix element,
\begin{equation}\label{Eq.:C6v phase factors}
\blangle X_m\,\s\,\?\Hh\? X_n\,\s'\brangle
=\bigl[\hat{\boldsymbol{s}}\times\boldsymbol{\mathrm{d}}_{m,n}\bigr]_{\s\s'}\,\frac{2}{3}i\lR\,,
\end{equation}
where, $\boldsymbol{\mathrm{d}}_{m,n}=\overrightarrow{mn\ }\bigl/|\overrightarrow{mn\ }|$ is the unit vector in the horizontal ($xy$) plane pointing from lattice site $n$ to nearest neighbor site $m$ and $\s\neq\s'$. So the general $C_{6v}$ invariant SOC Hamiltonian based on $\pi$-states, time-reversal and translational invariance reads,
\begin{align}\label{Eq.:C6v-SOC-Hamiltonian}
&\mathcal{H}_{C_{6v}}=
\frac{i\lI}{3\sqrt{3}}\sum\limits_\sigma\sum\limits_{\llangle m,n\rrangle}
\nu_{m,n}^{\phantom\dagger}\,\bigl[\hat{s}_z\bigr]_{\sigma\sigma}\,\?X_m\,\s\brangle\,\blangle X_n\,\s\?\nonumber\\
&+\frac{2\lH}{3}\sum\limits_{\s\neq\s'}\sum\limits_{\llangle m,n\rrangle} \bigl[i\hat{\boldsymbol{s}}\times\boldsymbol{\mathrm{d}}_{m,n}\bigr]_{\s\s'}\,\?X_m\,\s\brangle\,\blangle X_n\,\s'\?\nonumber\\
&+\frac{2i\lR}{3}\sum\limits_{\s\neq\s'}\sum\limits_{\langle m,n\rangle} \bigl[\hat{\boldsymbol{s}}\times\boldsymbol{\mathrm{d}}_{m,n}\bigr]_{\s\s'}\,\?X_m\,\s\brangle\,\blangle X_n\,\s'\?\,.
\end{align}
The first and the last term in $\mathcal{H}_{C_{6v}}$ are the well known SOC terms from the seminal papers of Kane and Mele\cite{Kane_Mele_1_2005:PRL,Kane_Mele_2_2005:PRL}. However, the staggered potential $\Delta$ added and considered by them is in fact not compatible with the $C_{6v}$ symmetry, but rather the $C_{3v}$ one discussed later.
What is more striking is the presence of the second---$\lH$ SOC term---which seems to be generally overseen by the community. Readers can easily convince themselves that there are not enough symmetries in $C_{6v}$ that can cancel its appearance in $\mathcal{H}_{C_{6v}}$. Indeed, to map the real matrix element $\langle A_3\ua|\Hh|A_2\da \rangle\sim\lH$, Eq.~(\ref{Eq.:PIA_D3d}), to ``$\pm$ itself'' within the pool of $C_{6v}$ symmetries, one can use respectively the vertical, $\syz$, and dihedral, $\sxz$, reflections---both flip spins---and accompany them by the rotation $\mathcal{R}^{\hat{z}}_\pi$, see Figs.~\ref{fig:hex_sitelabel_coordinate}~and~\ref{fig:d6h_subgroups}.
Since $\mathcal{R}^{\hat{z}}_\pi=\syz\circ\sxz$, the composition $\mathcal{S}=\mathcal{R}^{\hat{z}}_\pi\circ\sxz\circ\syz$ is the identity in the orbital and also in the spin space and hence $\mathcal{S}:|A_i\,\s\rangle\rightarrow|A_i\,\s\rangle$.
So at the end $\langle A_3\ua|\Hh|A_2\da\rangle=+\langle A_3\ua|\Hh|A_2\da\rangle$ what gives no constraint on $\lH$.

The first two terms in $\mathcal{H}_{C_{6v}}$, Eq.~(\ref{Eq.:C6v-SOC-Hamiltonian})---as we have discussed earlier---are not causing SOC splitting of the electronic band structure. The band SOC splitting is solely due to the space-inversion breaking term---Rashba SOC $\lR$. One can anticipate this fact also from the generally valid argument of Bychkov and Rashba\cite{Bychkov-Rashba1983:JETP-Letters}. They showed that a SOC induced band spin-splitting would appear in systems with a single high-symmetry (at least three-fold) axis, in our case the transverse $z$-axis, and an invariant vector along this axis, in our case the transverse electric field or the outward direction from the surface, what is exactly the case of $C_{6v}$ group and its subgroups.

In what follows we transform the $C_{6v}$-invariant SOC Hamiltonian into the Bloch form, $\mathcal{H}_{C_{6v}}=\sum_{\mathbf{q}}\mathcal{H}_{C_{6v}}(\mathbf{q})$.
The first two terms entering $\mathcal{H}_{C_{6v}}(\mathbf{q})$ can be compactly expressed in terms of Eq.~(\ref{Eq.:D3d-SOC-Hamiltonian-Bloch Form}), therefore we write here explicitly only the Rashba one
\begin{widetext}
\begin{align}
\mathcal{H}_{\mathrm{R}}(\mathbf{q})=i\lR\sum\limits_{\s,\s'}\sum\limits_{X,X'}
&\left\{
\bigl[\hat{\s}_{+}\bigr]_{XX'}
\left(f_{\mathrm{R}}(\mathbf{q})\,\bigl[\hat{s}_{+}\bigr]_{\s\s'} + \overline{f_{\mathrm{R}}(-\mathbf{q})}\,\bigl[\hat{s}_{-}\bigr]_{\s\s'}\right)\right. \nonumber \\
&\hspace{-1.7mm}-\left. \bigl[\hat{\s}_{-}\bigr]_{XX'}
\left(\overline{f_{\mathrm{R}}(\mathbf{q})}\,\bigl[\hat{s}_{-}\bigr]_{\s\s'} + f_{\mathrm{R}}(-\mathbf{q})\,\bigl[\hat{s}_{+}\bigr]_{\s\s'}\right)\right\}\,\?X_{\mathbf{q}}\,\s\brangle\blangle X_{\mathbf{q}}^\prime\,\s'\?\,,
\end{align}
\end{widetext}
i.e.~$\mathcal{H}_{C_{6v}}(\mathbf{q})=\mathcal{H}_{D_{3d}}(\mathbf{q})+\mathcal{H}_{\mathrm{R}}(\mathbf{q})$. The Rashba SOC structural function is given as follows,
\begin{equation}\label{Eq:structural-R-funcion}
f_{\mathrm{R}}(\mathbf{q})=\frac{2}{3}\Bigl\{1+\mathrm{e}^{-i\frac{2\pi}{3}}\mathrm{e}^{-i\mathbf{q}\cdot\mathbf{R}_3}+
\mathrm{e}^{i\frac{2\pi}{3}}\mathrm{e}^{i\mathbf{q}\cdot\mathbf{R}_2}\Bigr\}\,,
\end{equation}
and the sublattice raising/lowering operators are defined by $\hat{\s}_{\pm}=\tfrac{1}{2}(\hat{\s}_x\pm i\hat{\s}_y)$. In our sublattice convention we particulary have ~$\bigl[\hat{\s}_{+}\bigr]_{AB}=1=\bigl[\hat{\s}_{-}\bigr]_{BA}$ and $\bigl[\hat{\s}_{+}\bigr]_{BA}=0=\bigl[\hat{\s}_{-}\bigr]_{AB}$.
It is worth to mention that there does not exist a simple SOC field $\mathbf{n}(\mathbf{q})\cdot\hat{\boldsymbol{s}}$ such that the operator $\hat{\s}_0\,[\mathbf{n}(\mathbf{q})\cdot\hat{\boldsymbol{s}}]$ commutes with $\mathcal{H}_{C_{6v}}(\mathbf{q})$.

\begin{widetext}
\begin{figure*}
  \includegraphics[width=2.\columnwidth]{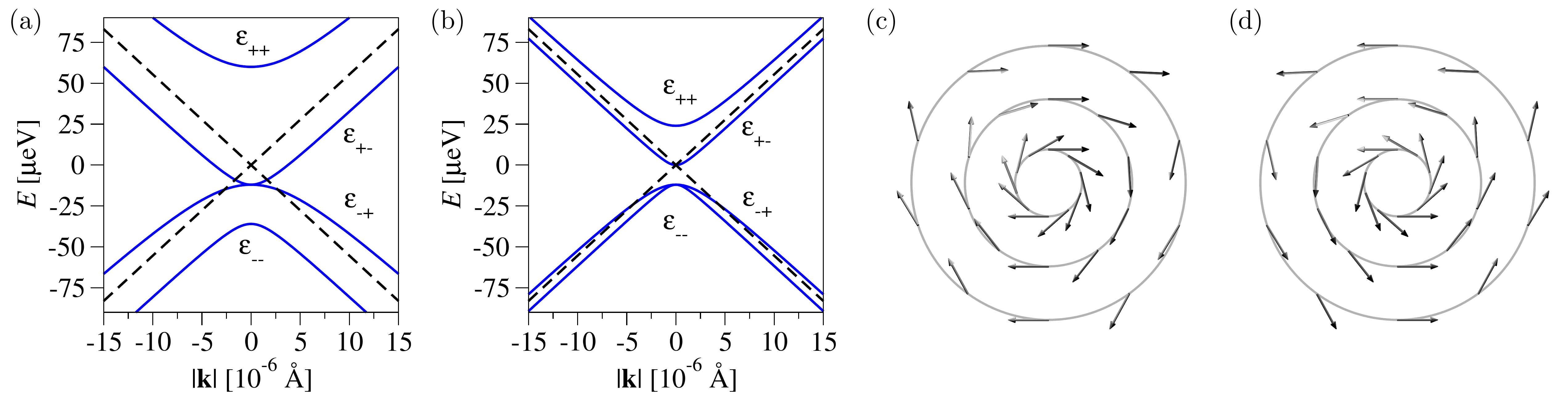}
\caption{(Color online)
Electronic band structure around the K-point in the presence of $C_{6v}$ invariant SOC Hamiltonian for $t=2.6$\,eV, $\lIA= 12\,\mu$eV and $\lH=0$ and (a) $\lR = 24\,\mu$eV and (b) $\lR = 6\,\mu$eV, respectively.
Panels (c) and (d) display the (in-plane) spin-orbit field $\mathbf{n(q)}$ around K-point for $\varepsilon_{--}$ and $\varepsilon_{-+}$ bands, respectively.
Circles radii correspond to 10\%, 30\% and 50\% of $\overline{\mathrm{KM}}$-distance and the model parameters are the same as used at panel (a).
Non-zero $\lH$ would add a k-dependent out of plane component of $\mathbf{n(q)}$ and slightly renormalize the Fermi velocity $v_F$.
\label{fig:c6v_band_vs_spin}}
\end{figure*}
\end{widetext}

The low energy expansion $f_{\mathrm{R}}(\tau\mathbf{K}+\mathbf{k})$ to the first order in $\mathbf{k}$ can be summarized by,
\begin{equation}
f_{\mathrm{R}}(\tau\mathbf{K}+\mathbf{k})=
\begin{cases}
2+i\frac{2}{\sqrt{3}}k_y a_{\mathrm{L}} & \mbox{for}\ +\mathbf{K}\,,\\
-\frac{1}{\sqrt{3}}k_x a_{\mathrm{L}}-i\frac{1}{\sqrt{3}}k_y a_{\mathrm{L}} & \mbox{for}\ -\mathbf{K}\,.
\end{cases}
\end{equation}
Since Rashba SOC is off-diagonal in spin and sublattice spaces, it is common to approximate $f_{\mathrm{R}}(\tau\mathbf{K}+\mathbf{k})$ by $f_{\mathrm{R}}(\tau\mathbf{K})$. Doing so we get the effective $C_{6v}$-invariant low energy Hamiltonian,
\begin{align}\label{Eq.:Effective-C6v}
\mathcal{H}&_{\mathrm{eff}}(\tau\mathbf{K}+\mathbf{k})=\hbar v_F\,\bigl(\tau k_x\hat{\s}_x-k_y\hat{\s}_y\bigr)\hat{s}_0+\tau\lI\hat{\s}_z\hat{s}_z+\nonumber\\
&+\lH\hat{\s}_z(k_x \hat{s}_y - k_y \hat{s}_x)a_{\mathrm{L}}-\lR(\tau\hat{\s}_x \hat{s}_y+\hat{\s}_y \hat{s}_x)\,.
\end{align}
whose eigenspectrum labeled by $n=\pm$ and $n^\prime=\pm$ reads,
\begin{align}\label{Eq:EigenValuesC6v}
\varepsilon_{n,n^\prime}(&\tau \mathbf{K}+\mathbf{k})=n^\prime\lR+\\
&+n\sqrt{\left(\lI+n^\prime\lR\right)^2+(\hbar^2v_F^2+\lH^2 a_{\mathrm{L}}^2)(k_x^2+k_y^2)}\,.\nonumber
\end{align}
Similarly as before, $\mathbf{k}$ is the quasi-momentum measured with respect to the given $\tau\mathbf{K}$-valley, $n=\pm$ stands for the conduction and valence bands, respectively, and the index $n^\prime=\pm$ stands for the spin polarization.
The spin expectation value---spin-orbit field $\mathbf{n(q)}$---at the given $\mathbf{q}$ and the band indices $n$ and $n'$ can be computed from the normalized eigenstates $|\mathbf{q},n,n^\prime\rangle$ via $\mathbf{n(q)}=\langle \mathbf{q},n,n^\prime|\hat{\boldsymbol{s}}|\mathbf{q},n,n^\prime\rangle$. The general formula is too complex and therefore we present only a result for the low energy eigenstates $|\tau\mathbf{K}+\mathbf{k},n,n^\prime\rangle$
of Eq.~(\ref{Eq.:Effective-C6v}) around the $\tau$K-valley:
 \begin{align}
 \left(
 \begin{array}{c}
 \langle\hat{s}_x\rangle \\
 \langle\hat{s}_y\rangle \\
 \langle\hat{s}_z\rangle
 \end{array}
 \right)=
 \frac{n^\prime}{N_{n,n^\prime}(k)}
 \left(
 \begin{array}{r}
 (\lI+\varepsilon_{n,n^\prime}) k_y \\
 -(\lI+\varepsilon_{n,n^\prime})k_x  \\
 \tau a_{\mathrm{L}}\lH k^2\phantom{\ \ \ }
 \end{array}
 \right)\,.
 \end{align}
Here $\varepsilon_{n,n^\prime}$ stands as a shorthand for eigenenergy $\varepsilon_{n,n^\prime}(\tau\mathbf{K}+\mathbf{k})$, see Eq.~(\ref{Eq:EigenValuesC6v}), $k^2\equiv k_x^2+k_y^2$, and $N_{n,n^\prime}(k)=\sqrt{(\lI+\varepsilon_{n,n^\prime})^2k^2+a_{\mathrm{L}}^2\lH^2 k^4}$.
For the visualization of the band structure and the spin-orbit field texture
see Fig.~\ref{fig:c6v_band_vs_spin}. It is worth to emphasize that at the Dirac points the two eigenvalues out of four become always degenerate. For example, for $\lI>\lR>0$ we have $\varepsilon_{-,-}(\tau\mathbf{K})=\varepsilon_{-,+}(\tau\mathbf{K})=-\lI$, and $\varepsilon_{+,\mp}(\tau\mathbf{K})=\lI\mp2\lR$ and the spectrum possesses the SOC induced gap with value $2(\lI-\lR)$.
For $\lR>\lI>0$ we have $\varepsilon_{-,+}(\tau\mathbf{K})=\varepsilon_{+,-}(\tau\mathbf{K})=-\lI$ and the spectral gap closes while
$\varepsilon_{-,-}(\tau\mathbf{K})=-2\lR+\lI$ and $\varepsilon_{+,+}(\tau\mathbf{K})=2\lR+\lI$. The case $\lI=\lR>0$ is critical, the spectrum changes
from gaped to gapless and we have a triple degeneracy $\varepsilon_{-,-}(\tau\mathbf{K})=\varepsilon_{-,+}(\tau\mathbf{K})=\varepsilon_{+,-}(\tau\mathbf{K})=-\lI$.

\subsubsection{$C_{3v}$-case: sublattice resolved $\lI$'s and $\lH$'s couplings, and $\lR$}\label{subsec:global C3v}
The point group $C_{3v}=\{E,2C^{\hat{z}}_{3},3\Sp_v\}$ is a subgroup of all three structural groups we have discussed earlier. For example, compared to the previous $C_{6v}$ case, the point group $C_{3v}$ lacks all the symmetries interchanging the sublattices. Hence the translationally invariant SOC Hamiltonian based on $\pi$-orbitals with $C_{3v}$ and time-reversal symmetries can be derived from the Hamiltonian $\mathcal{H}_{C_{6v}}$, Eq.~(\ref{Eq.:C6v-SOC-Hamiltonian}), inducing the next nearest neighbor SOC hoppings $i\lI$ and $\lH$ sublattice dependent, i.e.,~$i\lI\rightarrow\{i\lIA,i\lIB\}$ and $\lH\rightarrow\{\lHA,\lHB\}$:
\begin{align}\label{Eq.:C3v-SOC-Hamiltonian}
&\mathcal{H}_{C_{3v}}=
\frac{i\lIA}{3\sqrt{3}}\sum\limits_\sigma\sum\limits_{\llangle m,n\rrangle}
\nu_{m,n}^{\phantom\dagger}\,\bigl[\hat{s}_z\bigr]_{\sigma\sigma}\,\?A_m\,\s\brangle\,\blangle A_n\,\s\?\nonumber\\
&+\frac{i\lIB}{3\sqrt{3}}\sum\limits_\sigma\sum\limits_{\llangle m,n\rrangle}
\nu_{m,n}^{\phantom\dagger}\,\bigl[\hat{s}_z\bigr]_{\sigma\sigma}\,\?B_m\,\s\brangle\,\blangle B_n\,\s\?\nonumber\\
&+\frac{2\lHA}{3}\sum\limits_{\s\neq\s'}\sum\limits_{\llangle m,n\rrangle} \bigl[i\hat{\boldsymbol{s}}\times\boldsymbol{\mathrm{d}}_{m,n}\bigr]_{\s\s'}\,\?A_m\,\s\brangle\,\blangle A_n\,\s'\?\nonumber\\
&+\frac{2\lHB}{3}\sum\limits_{\s\neq\s'}\sum\limits_{\llangle m,n\rrangle} \bigl[i\hat{\boldsymbol{s}}\times\boldsymbol{\mathrm{d}}_{m,n}\bigr]_{\s\s'}\,\?B_m\,\s\brangle\,\blangle B_n\,\s'\?\nonumber\\
&+\frac{2i\lR}{3}\sum\limits_{\s\neq\s'}\sum\limits_{\langle m,n\rangle} \bigl[\hat{\boldsymbol{s}}\times\boldsymbol{\mathrm{d}}_{m,n}\bigr]_{\s\s'}\,\?X_m\,\s\brangle\,\blangle X_n\,\s'\?\,.
\end{align}
This Hamiltonian governs SOC effects in systems with broken sublattice symmetry (an effective staggered potential) and the fixed transverse direction (substrate or transverse electric field). Examples of such systems are semi-hydrogenated graphene\cite{Gmitra2013:PRL} (graphone), graphene/TMDC heterostructures\cite{Morpurgo2015:NComm,Gmitra2016:PRB(2)}, silicene on the substrate etc.

The Bloch form of the Hamiltonian $\mathcal{H}_{C_{3v}}$, Eq.~(\ref{Eq.:C3v-SOC-Hamiltonian}), is straightforward since all the structural functions---
$f_{\mathrm{I}}(\mathbf{q})$, $f_{\mathrm{P}}(\mathbf{q})$, $f_{\mathrm{R}}(\mathbf{q})$---were already given. Instead of that we fix the order of the Bloch basis $\{|A_{\mathbf{q}}\ua\rangle$, $|A_{\mathbf{q}}\da\rangle$, $|B_{\mathbf{q}}\ua\rangle$, $|B_{\mathbf{q}}\da\rangle\}$ and provide the low energy Hamiltonian around $\mathbf{q}=\tau\mathbf{K}+\mathbf{k}$, including the orbital term with the staggered potential, $\mathcal{H}_{\rm orb}(\tau\mathbf{K}+\mathbf{k})=\hbar v_F\,\bigl(\tau k_x\hat{\s}_x-k_y\hat{\s}_y\bigr)\hat{s}_0+\Delta\,\hat{\s}_z\hat{s}_0$, in the matrix form

\begin{widetext}
\begin{equation}\label{Eq.:Effective-C3v}
\mathcal{H}_{\mathrm{eff}}(\tau\mathbf{K}+\mathbf{k})=
\left(
\begin{array}{cccc}
\tau\lIA+\Delta & -\lHA(i k_x + k_y)a_{\mathrm{L}}  & \hbar v_F(\tau k_x+i k_y) & 2 i \lR\delta_{\tau\mathbf{K},+\mathbf{K}}\\
-\lHA(-i k_x + k_y)a_{\mathrm{L}}  & -\tau\lIA+\Delta& 2 i \lR\delta_{\tau\mathbf{K},-\mathbf{K}} & \hbar v_F(\tau k_x+i k_y)\\
\hbar v_F(\tau k_x-i k_y) & -2i \lR \delta_{\tau\mathbf{K},-\mathbf{K}} & -\tau\lIB-\Delta& \lHB(i k_x + k_y)a_{\mathrm{L}} \\
-2 i \lR\delta_{\tau\mathbf{K},+\mathbf{K}} & \hbar v_F(\tau k_x-i k_y) & \lHB(-i k_x + k_y)a_{\mathrm{L}}  & \tau\lIB-\Delta\\
\end{array}
\right)\,.
\end{equation}
\end{widetext}

\begin{widetext}
\begin{figure*}
\includegraphics[width=2.\columnwidth]{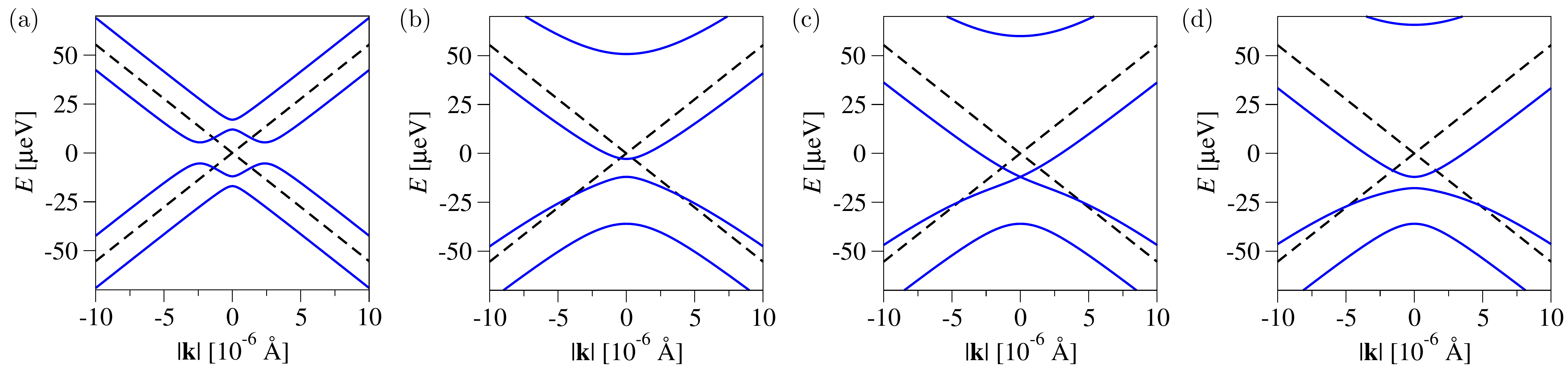}
\caption{(Color online)
Electronic band structure around the K-point in the presence of $C_{3v}$ invariant SOC Hamiltonian for $t=2.6$\,eV, staggered potential
$\Delta=$ and $\lIA= 12\,\mu$eV and (a) $\lIB=-\lIA$, $\lR = 6\,\mu$eV, (b) $\lIB=3\lIA$, $\lR =6\,\mu eV$, (c) $\lIB=3\lIA$, $\lR = \sqrt{2}\lIA$,
(d) $\lIB=3\lIA$, $\lR =20\,\mu eV$.
Panel (a) shows the lifting of band inversion (cf. Fig.~\ref{fig:d3h_lia12mlib_lb3lia}) for finite parameter $\lR$.
For cases without inversion, panel (b), $\lR$ can for special cases close the gap in the energy spectrum.
For comparison, the black dashed lines display the energy dispersion of the pristine graphene without SOC.
}
\end{figure*}
\end{widetext}

\section{Systems in absence of translational invariance---impurity induced SOC Hamiltonians}\label{sec:local}

In the forthcoming sections we discuss effective SOC Hamiltonians for hexagonal systems in the presence of locally chemisorbed impurities focusing on light ad-atoms and simple ad-molecules. The case of physisorbed heavy ad-elements is discussed in Ref.~[\onlinecite{Weeks2011:PRX,Pachoud2014:PRB}].
Since translational invariance is lost, the invariant expansion and decomposition into the irreps at high symmetry points in the Brillouin zone are not applicable. However, the tight-binding-like methodology based on the local atomic orbitals and their group symmetry properties allows us to treat this problem very naturally. We assume a dilute coverage by light adsorbates and hence it is enough to investigate local SOC effects due to a single chemisorbed impurity---cluster formation and interference SOC effects among nearby impurity centers are therefore not discussed.

The electronic structure of an adatom and host (in most cases graphene) and the underlying molecular dynamics determine mainly three stable binding positions: the \emph{hollow}, \emph{top}, and \emph{bridge} one.
Equivalently, we can distinguish those adatom configurations through their local point group symmetries: $C_{6v}$ for the hollow, $C_{3v}$ for the top, and $C_{2v}$ for the bridge one.
For simplicity we treat the chemisorbed ad-element as monovalent, i.e.,~it bonds via a single effective orbital that is invariant under the local point-group symmetries. This monovalency assumption seems to be crude, though experience shows that the effective single-orbital description works very well\cite{Gmitra2013:PRL,Irmer2015:PRB,Zollner2016:PRB,Frank2016:Cu}. However, an extension to the multi-orbital case is technically straightforward.

As already stated, we are interested in local effective SOC Hamiltonians in the presence of an impurity, that are invariant under the corresponding local point group symmetries. Those can be then added to the global translational invariant Hamiltonians of the host systems as discusses in the previous sections. Locality for us means hoppings up to the next nearest neighbors with respect to the adsorbed element. In what follows, we will label the adatom by $O$ and the corresponding atomic orbital by $|O\rangle$. Similarly, the adatom nearest neighbor sites and orbitals will be denoted by $Y_j$ and $|Y_j\rangle$, respectively, and the next nearest ones by $Z_j$ and $|Z_j\rangle$. The number of nearest and next nearest carbon neighbors may vary depending on the adsorption configuration---this is indicated by the subscript $j$.

From the orbital point of view the minimal tight-binding description of the adatom that chemisorbs with its nearest neighbors is given by
the Hamiltonian\cite{Robinson2008:PRL,Wehling2010:PRL,Gmitra2013:PRL,Irmer2015:PRB,Zollner2016:PRB,Frank2016:Cu} $\mathcal{H}_{\rm{orb}}$ which is defined
as,
\begin{align}\label{eq:Local Orb Hamiltonian}
\mathcal{H}_{\rm{orb}}= \omega&\sum\limits_{\s} \sum\limits_{\langle O, Y_j\rangle} |O\,\s \rangle \langle Y_j\,\s| + |Y_j\,\s\rangle \langle O\,\s|\nonumber\\
+ \varepsilon&\sum\limits_{\sigma} | O\,\s\rangle \langle O\,\s|\,.
\end{align}
The first term describes a hybridization $\omega$ between the ad-element and its nearest neighbors (summation over $\langle O, Y_j\rangle$) and the second represents the adatom's on-site energy. For the remaining orbitals we assume in the minimal-model scenario zero on-site contributions. The above orbital Hamiltonian is applicable to the hollow, top, and bridge configuration, respectively.

\subsection{Adatom in hollow position}
\begin{figure}[t]
 \includegraphics[width=0.5\columnwidth]{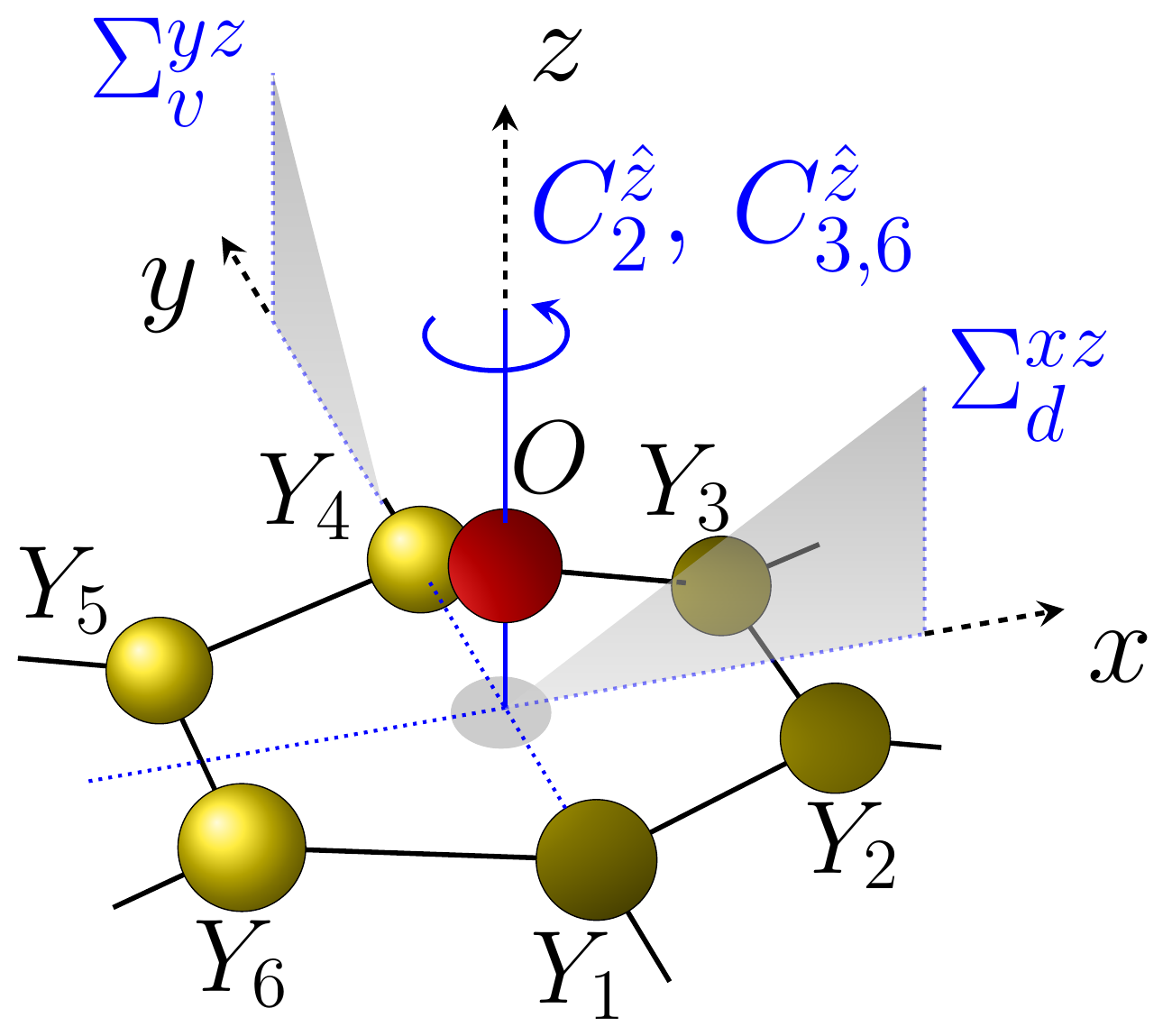}
 \caption{\label{Fig:hollow_symm}Adatom bonded in the hollow position: local point group symmetry $C_{6v}$---similar to graphene in the transverse
 external electric field, atom labeling convention, axes orientations and $C_{6v}$ group operations as discussed in the text.
 }
\end{figure}

\textit{Ab-initio} studies are unveiling that light metallic adatoms~\cite{Chan2008:PRB} from groups I-III and also heavy transition metals~\cite{Chan2008:PRB,Weeks2011:PRX,Mao:JOPCondMatt2008} favor to adsorb above the centers of graphene hexagons, i.e.~at the hollow positions. The same is true for light ad-molecules like NH$_3$, H$_2$O, NO$_2$.\cite{Leenaerts:PRB2008} The situation is schematically shown in Fig.~\ref{Fig:hollow_symm}. The central ad-element $O$ has six nearest carbon neighbors $Y_j$ and
since the out-of-plane position of the adatom fixes the orientation of the perpendicular $\hat{z}$, axis the structure is locally described by the point group $C_{6v}$. We will focus on a SOC Hamiltonian including the adatom orbital $|O\rangle$ and the $\pi$-state carbon orbitals $|Y_j\rangle$
of its direct nearest neighbors only. We first discuss the SOC mediated hoppings among $Y$'s sites and then we account for hoppings between the adatom orbital $|O\rangle$ and its six neighboring orbitals $|Y_j\rangle$. Since the translational symmetry is lost, we avoid using attributes like intrinsic, Rashba, and so on for the local SOC mediated hoppings. Instead we use the full taxonomy \emph{spin-conserving (next) nearest neighbor hopping}, $\Lambda^{\mathrm{(n)n}}_\mathrm{c}$, and \emph{spin-flipping (next) nearest neighbor hopping}, $\Lambda^{\mathrm{(n)n}}_\mathrm{f}$, respectively, reserving for the local SOC capital $\Lambda$.

The translationally invariant SOC Hamiltonian with $C_{6v}$ symmetry was discussed in the preceding section, Eq.~(\ref{Eq.:C6v-SOC-Hamiltonian}). Making it local, the global terms---$i\lI$, $\lH$ and $i\lR$---can not diminish. They would be respectively recast into their local analogs---$i\Lambda^{\mathrm{nn}}_\mathrm{c}$, $\Lambda^{\mathrm{nn}}_\mathrm{f}$ and $i\Lambda^{\mathrm{n}}_\mathrm{f}$. Since all the $Y$'s sites are equivalent, there are not sublattice resolved partners of those $\La$'s.
In addition, the lack of the translational invariance allows now also the purely imaginary spin-conserving nearest neighbor hopping $i\Lambda^{\mathrm{n}}_\mathrm{c}\simeq\langle Y_{i}\,\s|\Hh|Y_{i+1}\,\s\rangle$; see \emph{no-go arguments} of section~\ref{sec:no-go arguments}. So finally, there are four independent SOC mediated hoppings among the $Y$'s sites which---in analogy with the former analysis---can be defined as follows:
\begin{subequations}
\begin{align}
  i\La_{\rm c}^{\rm nn} &= \blangle Y_5 \ua \,|\, \hat{H}_{\rm so} \,|\, Y_3 \ua \brangle \,, \\
   \La_{\rm f}^{\rm nn} &=\blangle Y_5 \ua \,|\, \hat{H}_{\rm so} \,|\, Y_3 \da \brangle \,,\\
  i\La_{\rm f}^{\rm n}  &= \blangle Y_3 \ua \,|\, \hat{H}_{\rm so} \,|\, Y_2 \da \brangle \,, \\
  i\La_{\rm c}^{\rm n}  &= \blangle Y_2 \ua \,|\, \hat{H}_{\rm so} \,|\, Y_3 \ua \brangle \,;
\end{align}
\end{subequations}
for the labeling of atomic sites see Fig.~\ref{Fig:hollow_symm}. Here we no longer use the numerical prefactors $1\bigl/3\sqrt{3}$ and $2\bigl/3$, which were convenient for the low energy $k$-space expansions. The SOC mediated hoppings among the $Y$ sites at different configurations can be obtained by Eqs.~(\ref{Eq.:opposite spin SOC elements}), (\ref{Eq.:intr_graphene3}), and Eq.~(\ref{Eq.:C6v phase factors}). For $i\La_{\rm c}^{\rm n}$ we
have in analogy with Eq.~(\ref{Eq.:intr_graphene3}) the following identity which holds for any two nearest neighbors $Y_j$ and $Y_k$ of the adatom $O$:
\begin{align}
\blangle Y_j\,\s|\Hh|Y_k\,\s\brangle =\tilde{\nu}_{Y_j,Y_k}\,\bigl[\hat{s}_z\bigr]_{\s\s}\,i\La_{\rm c}^{\rm n}\,.
\end{align}
Here $\tilde{\nu}_{Y_j,Y_k}=+1(-1)$ if the hopping from the site $Y_k$ to $Y_j$ via a central adatom $O$ is counter clockwise (clockwise).

Next, we examine SOC mediated hoppings between the adatom orbital $|O\rangle$ and its neighbors $|Y_j\rangle$ along the hexagonal ring. For that it is enough
to look at matrix elements $\langle O\ua|\Hh|Y_1\ua\rangle$ and $\langle O\ua|\Hh|Y_1\da\rangle$, respectively. Assuming $|O\rangle$ is $C_{6v}$ and
time-reversal invariant---i.e.,~$\mathcal{S}|O\rangle=|O\rangle$ for any $\mathcal{S}\in C_{6v}$ and $\mathcal{T}|O\rangle=|O\rangle$ as would be the case of alkali metals---we can show that the first of the above matrix elements is identically zero and the second is purely imaginary. Particulary,
\begin{align}
\blangle O&\ua|\Hh|Y_1\ua\brangle
\overset{(\ref{eq:r_reflection_yz})}{=}
\blangle -i\syz[O\da]|\Hh|-i\syz[Y_1\da] \brangle\nonumber\\
&\overset{(\ref{Eq.:Unitary-Action})}{=}
\blangle O\da|\Hh|Y_1\da \brangle
\overset{(\ref{Eq.:pure_imaginary2})}{=}
-\blangle O\ua|\Hh|Y_1\ua \brangle\,.
\label{eq:hollow_forbid_xint}
\end{align}
For the spin-flip hopping we get,
\begin{align}
\blangle O&\ua|\Hh|Y_1\da\brangle
\overset{(\ref{eq:r_reflection_yz})}{=}
\blangle -i\syz[O\da]|\Hh|-i\syz[Y_1\ua] \brangle\nonumber\\
&\overset{(\ref{Eq.:Unitary-Action})}{=}
\blangle O\da|\Hh|Y_1\ua \brangle
\overset{(\ref{Eq.:opposite spin SOC elements})}{=}
-\overline{\blangle O\ua|\Hh|Y_1\da \brangle}\,,
\label{eq:hollow_allowed_xR}
\end{align}
what allows us to define the SOC term,
\begin{equation}\label{eq:hollow_L^{On}_f}
i\La^{\mathrm{On}}_{\mathrm{f}}=\blangle O\ua|\Hh|Y_1\da\brangle\,.
\end{equation}
Equivalent couplings can be specified by reflections, rotations and time-reversal---e.g.,~by applying $\rotvi$ we get,
\begin{align}
\blangle O&\ua|\Hh|Y_2\da\brangle = \blangle e^{i\frac{\pi}{6}}\rotvi[O\ua]|\Hh|e^{-i\frac{\pi}{6}}\rotvi[Y_1\da] \brangle\nonumber\\
&\overset{(\ref{Eq.:Unitary-Action})}{=}e^{-i\frac{\pi}{3}}\blangle O\ua|\Hh|Y_1\da\brangle
\overset{(\ref{eq:hollow_L^{On}_f})}{=} e^{-i\frac{\pi}{3}}\, i\La_{\rm f}^{\rm On}\,,
\end{align}
and in general, for any $Y_j$ and $\s\neq\s'$ we have,
\begin{align}
\blangle O\,\s|\Hh|Y_j\,\s'\brangle =\bigl[\hat{\boldsymbol{s}}\times\boldsymbol{\mathrm{d}}_{O,Y_j}\bigr]_{\s\s'}\,\, i\La_{\rm f}^{\rm On}\,,
\end{align}
where the meaning of $\boldsymbol{\mathrm{d}}_{O,Y_j}$ is identical as before---a unit vector in $xy$-plane pointing from site $Y_j$ to $O$.
The local SOC Hamiltonian for the hollow position possesses five SOC terms---$i\La_{\rm c}^{\rm n}$, $i\La_{\rm c}^{\rm nn}$,
$i\La_{\rm f}^{\rm n}$, $\La_{\rm f}^{\rm nn}$, and $i\La^{\mathrm{On}}_{\mathrm{f}}$---and is given as follows:
\begin{align}\label{eq:hollow_SOC Hamiltonian}
&\mathcal{H}^{\mathrm{hol}}_{\rm{so}}=
i\La^{\rm n\phantom{n}}_{\rm{c}}\sum\limits_{\s} \sum\limits_{\langle Y_j,Y_k\rangle}
\tilde{\nu}_{Y_j,Y_k}^{\phantom\dagger}\,\bigl[\hat{s}_z\bigr]_{\sigma\sigma}\,\? Y_j\,\s\brangle \blangle Y_k\,\s\?\nonumber\\
&+
i\La^{\rm nn}_{\rm{c}}\sum\limits_{\s} \sum\limits_{\llangle Y_j,Y_k\rrangle}
\nu_{Y_j,Y_k}^{\phantom\dagger}\,\bigl[\hat{s}_z\bigr]_{\sigma\sigma}\,\? Y_j\,\s\brangle \blangle Y_k\,\s\?\nonumber\\
&+
i\La^{\rm n\phantom{n}}_{\rm{f}}\sum\limits_{\s\neq\s'}\sum\limits_{\langle Y_j,Y_k\rangle} \bigl[\hat{\boldsymbol{s}}\times\boldsymbol{\mathrm{d}}_{Y_j,Y_k}\bigr]_{\s\s'}\,\?Y_j\,\s\brangle\,\blangle Y_k\,\s'\?\\
&+
\La^{\rm nn}_{\rm{f}}\sum\limits_{\s\neq\s'}\sum\limits_{\llangle Y_j,Y_k\rrangle} \bigl[i\hat{\boldsymbol{s}}\times\boldsymbol{\mathrm{d}}_{Y_j,Y_k}\bigr]_{\s\s'}\,\?Y_j\,\s\brangle\,\blangle Y_k\,\s'\?\nonumber\\
&+
i\La^{\rm On}_{\rm{f}}\sum\limits_{\s\neq\s'}\sum\limits_{\langle O,Y_j\rangle} \bigl[\hat{\boldsymbol{s}}\times\boldsymbol{\mathrm{d}}_{O,Y_j}\bigr]_{\s\s'}\,\?O\,\s\brangle\,\blangle Y_j\,\s'\? +\mathrm{hc}\,.\nonumber
\end{align}
Again, the summation over the nearest and next nearest neighbors is specified by $\langle\,,\,\rangle$ and $\llangle\,,\,\rrangle$ brackets, respectively; for the atomic configurations that enter $\nu,\tilde{\nu}$ and $\boldsymbol{\mathrm{d}}$ see Fig.~\ref{Fig:hollow_symm}.

\subsection{Adatom in top-position}
\begin{figure}[t]
 \includegraphics[width=0.65\columnwidth]{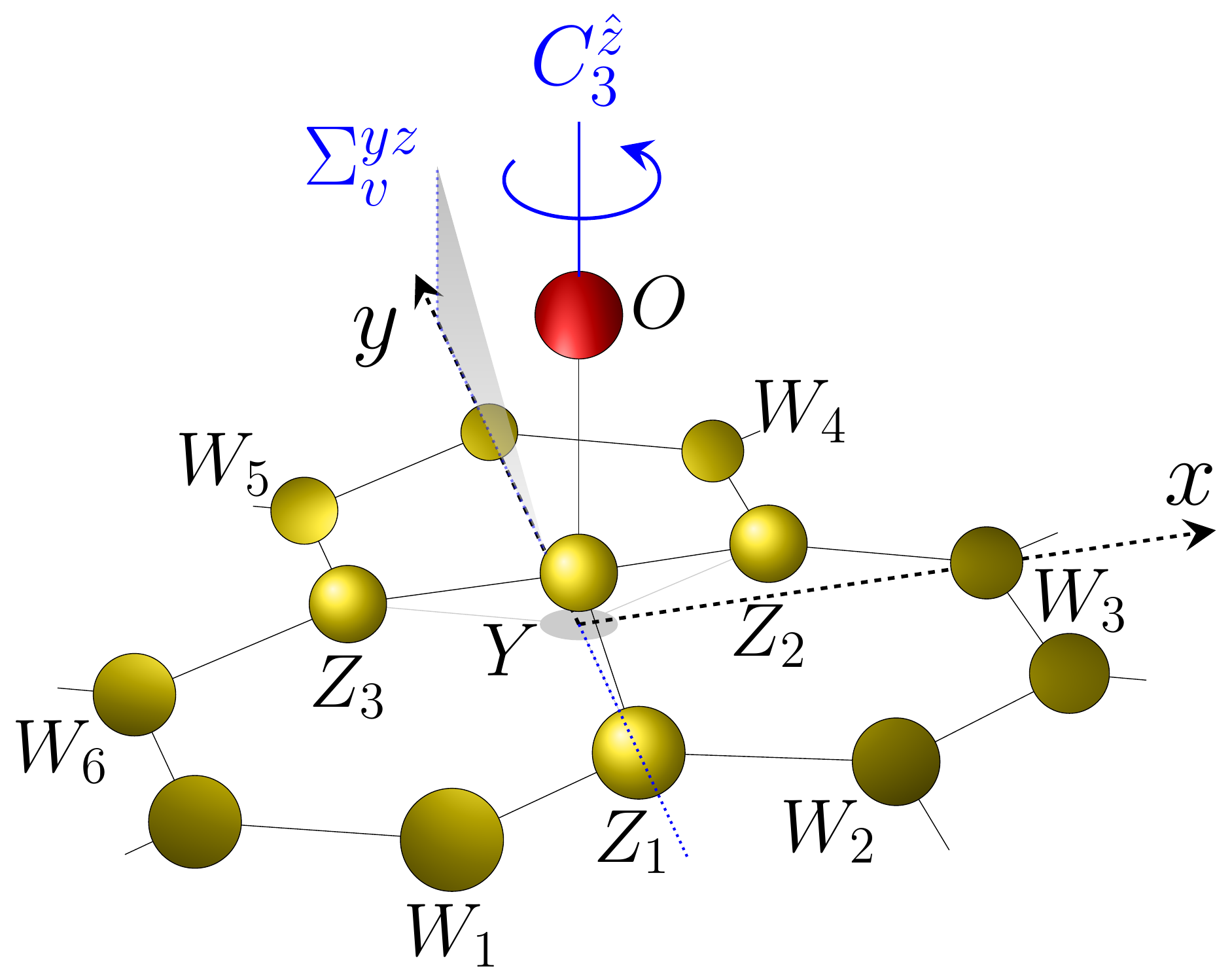}
 \caption{\label{fig:local_c3v}Adatom bonded in the top position with a center of symmetry on the bonding axis:
 the local point group symmetry $C_{3v}$, atom labeling convention, axes orientations and $C_{3v}$ group operations as discussed in the main text.
 }
\end{figure}

Adsorption in the top position seems to be favorable for light atoms like hydrogen\cite{Boukhvalov:PRB2008,Gmitra2013:PRL}, fluorine\cite{Wu:APL2008,Sahin:PRB2011,Irmer2015:PRB} and copper\cite{Wu:APL2009,Amft:JPhysCondMat2011,Frank2016:Cu}, the heavier gold atom\cite{Chan2008:PRB,Amft:JPhysCondMat2011}, and, for example, also the light ad-molecule methyl\cite{Zollner2016:PRB}.
The model configuration has a local $C_{3v}$ point group symmetry and is displayed in Fig.~\ref{fig:local_c3v}---an adatom $O$ binding on the top possesses one nearest $Y$ neighbor, three second nearest $Z$ neighbors, and six third nearest $W$ neighbors. To compare the global and local $C_{3v}$ cases which have different centers of symmetry we consider also mutual SOC hoppings implementing the third-nearest $W$ neighbors.

The local $C_{3v}$-invariant SOC Hamiltonian accounting for SOC mediated hoppings among the $Y$, $Z$, and $W$ carbon sites---the SOC hoppings connecting the adatom will be discussed later---can be naturally derived from the global $C_{3v}$ Hamiltonian, Eq.~(\ref{Eq.:C3v-SOC-Hamiltonian}). In analogy with the global $i\lIA$, $i\lIB$, $\lHA$, $\lHB$, and $i\lR$ couplings we correspondingly have,
\begin{subequations}\label{eq:local C3v SOC part1}
\begin{align}
  i\La_{\rm c}^{\rm YW} &= \blangle Y \ua \,|\, \hat{H}_{\rm so} \,|\, W_3 \ua \brangle   \,, \\
  i\La_{\rm c}^{\rm ZZ} &= \blangle Z_3 \ua \,|\, \hat{H}_{\rm so} \,|\, Z_2 \ua \brangle \,, \\
   \La_{\rm f}^{\rm YW} &= \blangle Y \ua \,|\, \hat{H}_{\rm so} \,|\, W_3 \da \brangle    \,, \\
   \La_{\rm f}^{\rm ZZ} &= \blangle Z_3 \ua \,|\, \hat{H}_{\rm so} \,|\, Z_2 \da \brangle  \,, \\
  i\La_{\rm f}^{\rm YZ} &=\blangle Y \ua \,|\, \hat{H}_{\rm so} \,|\, Z_1 \da \brangle    \,.
\end{align}
\end{subequations}
Here again the subscripts ``c'' and ``f'' stand for spin-conserving and spin-flipping hoppings, respectively and the superscripts made from $Y$, $Z$, and $W$ encode particular nearest or next nearest neighbor hoppings among the $Y$, $Z$, and $W$ carbon sites. For atomic configuration and labeling see Fig.~\ref{fig:local_c3v}.

Using Eq.~(\ref{Eq.:pure_imaginary}) we see that $i\La_{\rm c}^{\rm YW}$ and
$i\La_{\rm c}^{\rm ZZ}$---local analogs of $i\lIA$ and $i\lIB$---are purely imaginary. Similarly, substituting in Eq.~(\ref{Eq.:D3d pure real lH}) $A_2$ by $Z_2$ and $A_3$ by $Z_3$ unveils that $\La_{\rm f}^{\rm ZZ}$---a local analog of $\lHB$---is purely real. Readapting the argumentation used in Eq.~(\ref{Eq.:pure_imaginary_Rashba}) to the situation displayed at Fig.~\ref{fig:local_c3v} we get,
\begin{align}
\blangle Y&\ua\?\Hh\?Z_1\da\brangle=\blangle -i\syz[Y\da]\?\Hh\? -i\syz[Z_1\ua]\brangle\nonumber\\
&\overset{(\ref{Eq.:Unitary-Action})}{=}\blangle Y\da\?\Hh\? Z_1\ua\brangle
\overset{(\ref{Eq.:opposite spin SOC elements})}{=}-\overline{\blangle Y\ua\?\Hh\? Z_1\da\brangle}\,,
\end{align}
what confirms that $ i\La_{\rm f}^{\rm YZ}$---a local analog of $i\lR$---is purely imaginary.

What differs from the global $C_{3v}$ case is the spin-flip coupling $\La_{\rm f}^{\rm YW}$---an analog of $\lHA$. Since now the sites $Y$ and $W$ are not interchangeable we cannot use the argument analogous to Eq.~(\ref{Eq.:D3d pure real lH}) and hence $\La_{\rm f}^{\rm YW}$ is in general complex. This slightly affects the former phase-factor formula, Eq.~(\ref{Eq.:D3d phase factors}), which now reads,
\begin{align}
\blangle Y\,&\s\,\?\Hh\? W_j\,\s'\brangle=\\
&=\bigl[i\hat{\boldsymbol{s}}\times\boldsymbol{\mathrm{d}}_{Y,W_j}\bigr]_{\s\s'}\,\Bigl[\mathrm{Re}\bigl(\La_{\rm f}^{\rm YW}\bigr)+i\nu_{Y,W_j}\,\mathrm{Im}\bigl(\La_{\rm f}^{\rm YW}\bigr)\Bigr]\,.\nonumber
\end{align}
The meaning of $\boldsymbol{\mathrm{d}}_{Y,W_j}$ and $\nu_{Y,W_j}$ stays the same as before. The spin-orbit mediated hoppings among the carbon atoms in the vicinity of the impurity site $O$ are now recapped.

In what follows we discuss the SOC mediated hoppings $\langle O\,\s|\Hh|Y\,\s'\rangle$ and $\langle O\,\s|\Hh|Z_i\,\s'\rangle$ that couple directly to the adatom orbital $|O\rangle$---assuming it is $C_{3v}$ and time-reversal invariant.
Repeating the discussion at the end of previous section, see Eqs.~(\ref{eq:hollow_forbid_xint})~and~(\ref{eq:hollow_allowed_xR}), we immediately get
\begin{align}
i\La^{\mathrm{OZ}}_{\mathrm{c}}&=\blangle O\ua|\Hh|Z_1\ua\brangle\equiv 0\,,\\
i\La^{\mathrm{OZ}}_{\mathrm{f}}&=\blangle O\ua|\Hh|Z_1\da\brangle\neq 0\,.\label{eq:local C3v SOC part2}
\end{align}
To show that $\langle O\ua|\Hh|Y\ua\rangle$ and $\langle O\ua|\Hh|Y\da\rangle$ are zero one can proceed as follows: for the first term we have,
\begin{align}
\blangle O\ua\?\Hh\?Y\ua\brangle
&\overset{(\ref{Eq.:pure_imaginary2})}{=}-\blangle O\da\?\Hh\?Y\da\brangle\nonumber\\
&\overset{(\ref{eq:r_reflection_yz})}{=}-\blangle -i\syz[O\ua]\?\Hh\?-i\syz[Y\ua]\brangle\nonumber\\
&\overset{(\ref{Eq.:Unitary-Action})}{=}-\blangle O\ua\?\Hh\?Y\ua\brangle\,,
\end{align}
what implies that $\langle O\ua|\Hh|Y\ua\rangle\equiv 0$. To show that $\langle O\ua|\Hh|Y\da\rangle$ is zero we apply rotations $\mathcal{R}^{\hat{z}}_{\pm\frac{2\pi}{3}}\in C_{3v}$, then
\begin{align}
\blangle O\ua\?\Hh\?Y\da\brangle
&\overset{(\ref{eq:r_rotation})}{=}\blangle e^{\pm i\frac{\pi}{3}}\mathcal{R}^{\hat{z}}_{\pm\frac{2\pi}{3}}[O\ua]
\?\Hh\?
e^{\mp i \frac{\pi}{3}}\mathcal{R}^{\hat{z}}_{\pm\frac{2\pi}{3}}[Y\da]\brangle\nonumber\\
&\overset{(\ref{Eq.:Unitary-Action})}{=}
e^{\mp i \frac{2\pi}{3}}\,\blangle O\ua\?\Hh\?Y\da\brangle\,.
\end{align}
The above relation can be fulfilled only by zero, therefore $\langle O\ua|\Hh|Y\da\rangle=0$.

Summarizing Eqs.~(\ref{eq:local C3v SOC part1})~and~(\ref{eq:local C3v SOC part2}), we have in total six spin-orbit couplings---four purely imaginary $i\La_{\rm c}^{\rm YW}$, $i\La_{\rm c}^{\rm ZZ}$, $i\La_{\rm f}^{\rm YZ}$, $i\La_{\rm f}^{\rm OZ}$, one purely real
$\La_{\rm f}^{\rm ZZ}$, and one in general complex $\La_{\rm f}^{\rm YW}$. The local SOC Hamiltonian with $C_{3v}$ symmetry that corresponds to the impurity in the top position reads
\begin{widetext}
\begin{align}\label{eq:top_SOC Hamiltonian}
\mathcal{H}^{\mathrm{top}}_{\rm{so}}&=
i\La^{\rm YW}_{\rm{c}}\sum\limits_{\s} \sum\limits_{\llangle Y,W_j\rrangle}
\nu_{Y,W_j}^{\phantom\dagger}\,\bigl[\hat{s}_z\bigr]_{\sigma\sigma}\,\? Y\,\s\brangle \blangle W_j\,\s\?+\mathrm{hc}\nonumber\\
&+
i\La^{\rm ZZ}_{\rm{c}}\sum\limits_{\s} \sum\limits_{\llangle Z_j,Z_k\rrangle}
\nu_{Z_j,Z_k}^{\phantom\dagger}\,\bigl[\hat{s}_z\bigr]_{\sigma\sigma}\,\? Z_j\,\s\brangle \blangle Z_k\,\s\?\nonumber\\
&+
\sum\limits_{\s\neq\s'}\sum\limits_{\llangle Y,W_j\rrangle}
\bigl[i\hat{\boldsymbol{s}}\times\boldsymbol{\mathrm{d}}_{Y,W_j}\bigr]_{\s\s'}\,\Bigl[\mathrm{Re}\bigl(\La_{\rm f}^{\rm YW}\bigr)+i\nu_{Y,W_j}\,\mathrm{Im}\bigl(\La_{\rm f}^{\rm YW}\bigr)\Bigr]\,\?Y\,\s\brangle\,\blangle W_j\,\s'\?+\mathrm{hc}\nonumber\\
&+
\La^{\rm ZZ}_{\rm{f}}\sum\limits_{\s\neq\s'}\sum\limits_{\llangle Z_j,Z_k\rrangle} \bigl[i\hat{\boldsymbol{s}}\times\boldsymbol{\mathrm{d}}_{Z_j,Z_k}\bigr]_{\s\s'}\,\?Z_j\,\s\brangle\,\blangle Z_k\,\s'\?\nonumber\\
&+
i\La^{\rm YZ}_{\rm{f}}\sum\limits_{\s\neq\s'}\sum\limits_{\langle Y,Z_j\rangle} \bigl[\hat{\boldsymbol{s}}\times\boldsymbol{\mathrm{d}}_{Y,Z_j}\bigr]_{\s\s'}\,\?Y\,\s\brangle\,\blangle Z_j\,\s'\?+\mathrm{hc}\nonumber\\
&+
i\La^{\rm OZ}_{\rm{f}}\sum\limits_{\s\neq\s'}\sum\limits_{\langle O,Z_j\rangle} \bigl[\hat{\boldsymbol{s}}\times\boldsymbol{\mathrm{d}}_{O,Z_j}\bigr]_{\s\s'}\,\?O\,\s\brangle\,\blangle Z_j\,\s'\?+\mathrm{hc}\,.
\end{align}
\end{widetext}

\subsection{Adatom in bridge position}
\begin{figure}[t]
 \includegraphics[width=0.5\columnwidth]{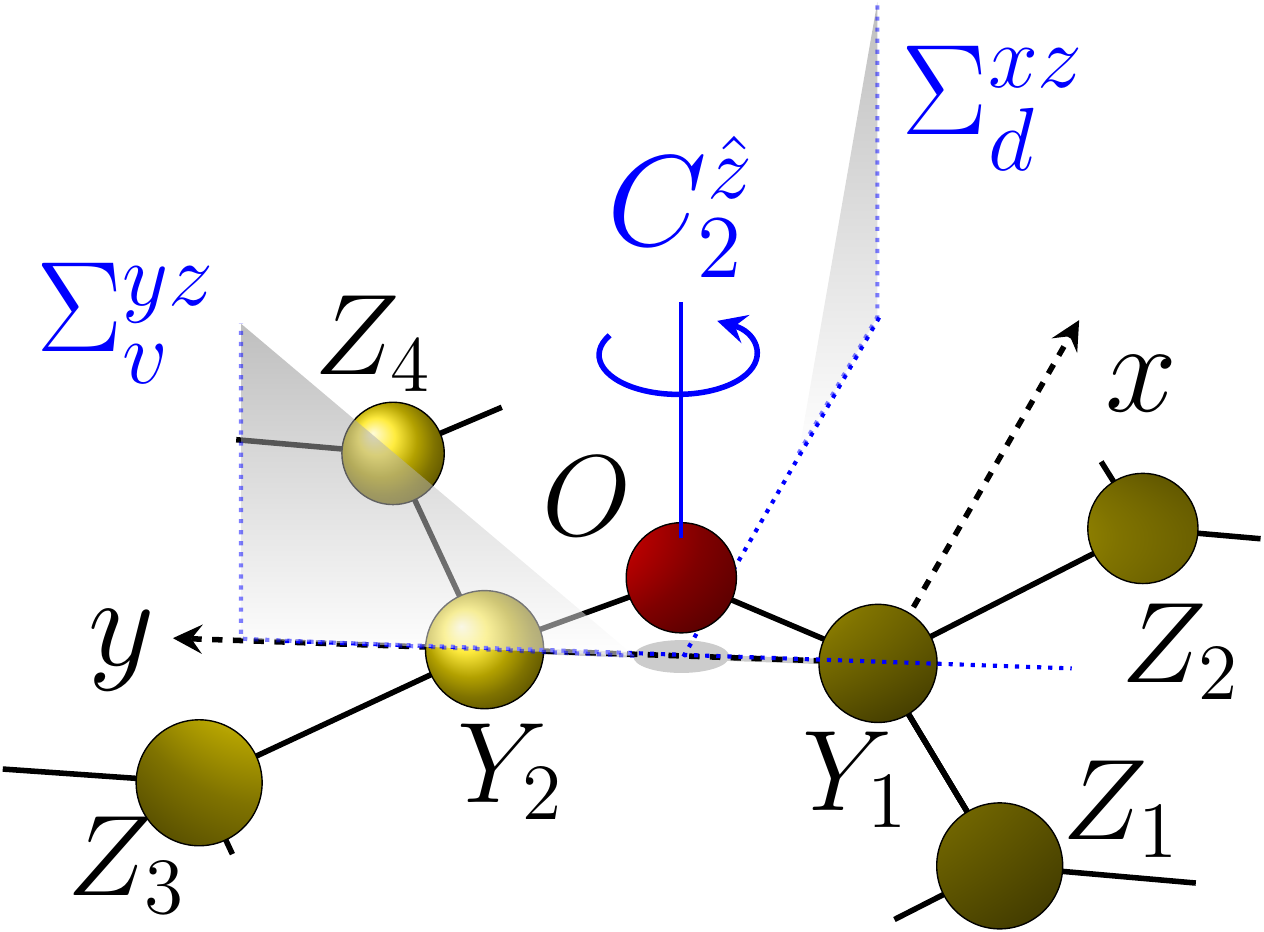}
 \caption{\label{Fig:bridge_symm} Adatom bonded in the bridge position with a center of symmetry on the vertical axis passing the adatom:
 the local point group symmetry $C_{2v}$, atom labeling convention, axes orientations and $C_{2v}$ group operations as discussed in the main text.
 }
\end{figure}

Oxygen and nitrogen are theoretically predicted to bond in the bridge position\cite{Wu:APL2008}. However, also for impurities in the top position like copper\cite{Wu:APL2009,Amft:JPhysCondMat2011,Frank2016:Cu} and
gold\cite{Chan2008:PRB,Amft:JPhysCondMat2011} the energy difference between the top and bridge configurations is relatively small and therefore their bridge realization becomes quite probable. Similarly, the light ad-molecules like CO, NO and NO$_2$ prefer to adsorb\cite{Leenaerts:PRB2008} equally-likely to the hollow and bridge positions. For those reasons we discuss in this section an effective SOC Hamiltonian that works for light ad-elements in the bridge configuration.
Particulary, by bridge we understand a configuration when the adatom $O$ splits a nearest neighbor bond between two---$Y_1$ and $Y_2$---carbon sites, see Fig.~\ref{Fig:bridge_symm}.
Such a structure possesses $C_{2v}$ point group symmetry which comprises two non-equivalent reflection planes $\sxz$ and $\syz$, and
$C_2$ rotation around the axis of their intersection; see Fig.~\ref{Fig:bridge_symm}.
As the order of this group is lower compared to the above cases we expect more SOC mediated matrix elements which in general would be complex-valued. Even within the approximation that keeps only nearest and next nearest neighbor hoppings among $O$, $Y$ and $Z$ sites, the effective SOC Hamiltonian contains eight hoppings. Three of them are spin-conserving (and hence purely imaginary) and the remaining five are spin-flipping,
\begin{subequations}
\begin{align}
  i\La_{\rm c}^{\rm YZ} &= \blangle Y_2 \ua \,|\, \hat{H}_{\rm so} \,|\, Z_4 \ua \brangle \,, \label{eq:br_lnc}\\
  i\La_{\rm c}^{\rm OZ} &= \blangle O \ua \,|\, \hat{H}_{\rm so} \,|\, Z_4 \ua \brangle \,, \label{eq:br_lxnnc}\\
  i\La_{\rm c}^{\rm ZZ} &= \blangle Z_1 \ua \,|\, \hat{H}_{\rm so} \,|\, Z_2 \ua \brangle \,, \label{eq:br_lnnc}\\
  i\La_{\rm f}^{\rm OY} &= \blangle O \ua \,|\, \hat{H}_{\rm so} \,|\, Y_1 \da \brangle \,, \label{eq:br_lxnf}\\
  i\La_{\rm f}^{\rm YY} &= \blangle Y_1 \ua \,|\, \hat{H}_{\rm so} \,|\, Y_2 \da \brangle \,, \label{eq:br_lnabf}\\
  {\rm Re}(\La_{\rm f}^{\rm YZ}) + {\rm i}\,{\rm Im}(\La_{\rm f}^{\rm YZ}) &= \blangle Y_2 \ua \,|\, \hat{H}_{\rm so} \,|\, Z_4 \da \brangle \,, \label{eq:br_lnf}\\
  {\rm Re}(\La_{\rm f}^{\rm OZ}) + {\rm i}\,{\rm Im}(\La_{\rm f}^{\rm OZ}) &= \blangle O \ua \,|\, \hat{H}_{\rm so} \,|\, Z_4 \da \brangle \,, \label{eq:br_lxnnf}\\
  \La_{\rm f}^{\rm ZZ} &= \blangle Z_2 \ua \,|\, \hat{H}_{\rm so} \,|\, Z_1 \da \brangle \,. \label{eq:br_lnnf}
\end{align}
\end{subequations}
Let us shortly comment on three of the above emerging couplings---Eqs.~(\ref{eq:br_lnc}),~(\ref{eq:br_lnf})~and~(\ref{eq:br_lxnnf}).
The absence of the translational invariance allows spin-conserving hopping $i\La_{\rm c}^{\rm YZ}$ between the nearest neighbor sites $Y$ and
$Z$---similar coupling was encountered in the local $C_{6v}$ case for the adatom in hollow position. For the same reason there
are the spin-flip hoppings $\La_{\rm f}^{\rm YZ}$---among the nearest neighbors---and $\La_{\rm f}^{\rm OZ}$---among the next nearest neighbors and both are complex-valued in general.

Altogether, we can write the local Hamiltonian for the local $C_{2v}$ symmetric structure in a closed form, with the help of the definitions that we introduced above, as follows:
\begin{widetext}
\begin{align}\label{eq:bridge_SOC Hamiltonian}
\mathcal{H}^{\mathrm{brid}}_{\rm{so}}&=
i\La^{\rm YZ}_{\rm{c}}\sum\limits_{\s} \sum\limits_{\langle Y_j,Z_k\rangle}
\nu_{O,Z_k}^{\phantom\dagger}\,\bigl[\hat{s}_z\bigr]_{\sigma\sigma}\,\? Y_j\,\s\brangle \blangle Z_k\,\s\?+\mathrm{hc}\nonumber\\
&+
i\La^{\rm OZ}_{\rm{c}}\sum\limits_{\s} \sum\limits_{\llangle O,Z_k\rrangle}
\nu_{O,Z_k}^{\phantom\dagger}\,\bigl[\hat{s}_z\bigr]_{\sigma\sigma}\,\? O\,\s\brangle \blangle Z_k\,\s\?+\mathrm{hc}\nonumber\\
&+
i\La^{\rm ZZ}_{\rm{c}}\sum\limits_{\s} \sum\limits_{\llangle Z_j,Z_k\rrangle}
\nu_{Z_j,Z_k}^{\phantom\dagger}\,\bigl[\hat{s}_z\bigr]_{\sigma\sigma}\,\? Z_j\,\s\brangle \blangle Z_k\,\s\?\nonumber\\
&+
i\La^{\rm OY}_{\rm{f}}\sum\limits_{\s\neq\s'}\sum\limits_{\langle O,Y_j\rangle} \bigl[\hat{\boldsymbol{s}}\times\boldsymbol{\mathrm{d}}_{O,Y_j}\bigr]_{\s\s'}\,\?O\,\s\brangle\,\blangle Y_j\,\s'\?+\mathrm{hc}\nonumber\\
&+
i\La^{\rm YY}_{\rm{f}}\sum\limits_{\s\neq\s'}
\bigl[\hat{\boldsymbol{s}}\times\boldsymbol{\mathrm{d}}_{Y_1,Y_2}\bigr]_{\s\s'}\,\?Y_1\,\s\brangle\,\blangle Y_2\,\s'\?+\mathrm{hc}\nonumber\\
&+
\sum\limits_{\s\neq\s'}\sum\limits_{\langle Y_j,Z_k\rangle}
\Bigl\{\nu_{O,Z_k}\bigl[i\hat{s}_y\bigr]_{\s\s'}\,\mathrm{Re}\bigl(\La_{\rm f}^{\rm YZ}\bigr)+i\,\mathrm{Im}\bigl(\La_{\rm f}^{\rm YZ}\bigr)\Bigr\}\,\mathrm{sgn}\bigl[\boldsymbol{\mathrm{d}}_{O,Y_j}\cdot\boldsymbol{\mathrm{d}}_{Y_1,Y_2}\bigr]\,\?Y_j\,\s\brangle\,\blangle Z_k\,\s'\?\nonumber+\mathrm{hc}\\
&+
\sum\limits_{\s\neq\s'}\sum\limits_{\llangle O,Z_j\rrangle}
\Bigl\{\nu_{O,Z_j}\bigl[i\hat{s}_y\bigr]_{\s\s'}\,\mathrm{Re}\bigl(\La_{\rm f}^{\rm OZ}\bigr)+i\,\mathrm{Im}\bigl(\La_{\rm f}^{\rm OZ}\bigr)\Bigr\}\,\mathrm{sgn}\bigl[\boldsymbol{\mathrm{d}}_{O,Z_j}\cdot\boldsymbol{\mathrm{d}}_{Y_1,Y_2}\bigr]\,\?O\,\s\brangle\,\blangle Z_j\,\s'\?+\mathrm{hc}\nonumber\\
&+
\La^{\rm ZZ}_{\rm{f}}\sum\limits_{\s\neq\s'}\sum\limits_{\llangle Z_j,Z_k\rrangle} \bigl[i\hat{\boldsymbol{s}}\times\boldsymbol{\mathrm{d}}_{Z_j,Z_k}\bigr]_{\s\s'}\,\?Z_j\,\s\brangle\,\blangle Z_k\,\s'\?\,.
\end{align}
\end{widetext}
In the first and sixth line, in which we are summing over the nearest neighbors $\langle Y_j, Z_k\rangle$, the symbol $\nu_{O,Z_k}$ has the following meaning;
it equals $1$ $(-1)$ if the path $Z_k\rightarrow Y_j$ after extension to the next nearest neighbor path $Z_k\rightarrow Y_j\rightarrow O$ becomes counter clockwise (clockwise).

\section{Final remarks $\&$ Conclusion}\label{sec:conclude}

\setlength{\tabcolsep}{6pt}
\begin{table*}[t]
\newcolumntype{C}{>{\centering\arraybackslash}p{2.4em}}
\newcolumntype{D}{>{\centering\arraybackslash}p{5.3em}}
\begin{tabular}{l|DDDDCCC}
\hline
Ad-element $\bigl/$ SOC [meV]$^{\phantom{^a}}_{\phantom{a_a}}$
& $\La^{\rm YW}_{\rm{c}}$  &  $\La^{\rm ZZ}_{\rm{c}}$ &  $\La^{\rm ZZ}_{\rm{f}}$ & $\La^{\rm YZ}_{\rm{f}}$ & $\La^{\rm YY}_{\rm{f}}$ & $\La^{\rm YZ}_{\rm{f}}$ & $\La^{\rm OZ\phantom{^{a^a}}}_{\rm{f}}$ \\
\hline\hline
 Hydrogen (top)\cite{Gmitra2013:PRL}& -0.04 & ---   & -0.51   & 0.22 & ---    & ---   & --- \\
Fluorine (top)\cite{Irmer2015:PRB} & --- & 0.64   & 4.87   & 7.47 & ---    & ---   & --- \\
Methyl (top)\cite{Zollner2016:PRB}   & -0.15 & 0.03    & -0.46   & 0.68 & ---    & ---   & --- \\
Copper (top)\cite{Frank2016:Cu}   & --- & 1.73    & 31.6   & 20.1 & ---    & ---   & --- \\
Copper (bridge)\cite{Frank2016:Cu}& --- & ---    & ---   & --- &  41.0   & -7.5   & 1.4+i$\cdot$8.4 \\
\hline
\end{tabular}
\caption{\label{Tab:all-data2}Summary of local SOC strengths for different ad-elements: hydrogen, fluorine, methyl and copper.\cite{Gmitra2013:PRL,Irmer2015:PRB,Zollner2016:PRB,Frank2016:Cu} Let us emphasize that compared to the referred manuscripts
and notation used therein---$\LIA$, $\LIB$, $\Lambda^\mathrm{B}_\mathrm{PIA}$, and $\LR$---we renamed and also properly rescaled their strengths to match the present convention.
A translation between the new and old notations is as follows: $\La^{\rm YW}_{\rm{c}}=\LIA/(3\sqrt{3})$, $\La^{\rm ZZ}_{\rm{c}}=\LIB/(3\sqrt{3})$, $\La^{\rm ZZ}_{\rm{f}}=2\Lambda^\mathrm{B}_\mathrm{PIA}/3$, and $\La^{\rm YZ}_{\rm{f}}=2\LR/3$.}
\end{table*}

As already noted, the lower the symmetry, the more SOC parameters enter the effective model Hamiltonians.
Before applying a particular model to spin-transport studies, two issues should be resolved. First, figure out the realistic strengths of SOC parameters and, second, reduce the number of the parameters as much as possible. For that, one should employ first-principles calculations together with physical intuition and common sense.

To describe our strategy, we start with \textit{ab-initio} calculations considering a large graphene supercell with one ad-element bonded in a given configuration. The larger the supercell, the weaker are the interactions among the periodic images, and the more representative the dilute coverage limit is realized. Analyzing local DOS and its atomic orbital decomposition, we directly test whether the system can be properly described by the adequate Hamiltonian model, i.e.~carbon $\pi$-orbitals and an effective ad-atom level. In all the cases yet analyzed---hydrogen\cite{Gmitra2013:PRL}, fluorine\cite{Irmer2015:PRB},
CH$_3$-group\cite{Zollner2016:PRB} and copper\cite{Frank2016:Cu} (both in top and bridge configurations)---the effective models with effective adatom orbitals work perfectly.
Fitting the spin-orbit induced band splittings would give us the strengths of the sought SOC parameters. The aim is to find a minimal set of best-fitting parameters to keep the model simple and simultaneously capture the main features in SOC induced band splittings. It might not be necessary to take into account all the symmetry-allowed coupling parameters. For that some intuition, experience and an input from the DFT are helpful, e.g.,~the possibility to turn off in first-principles calculations SOC interaction on the adatom, or shift away the Fermi level Bloch states composed from the (un)wanted atomic orbitals\cite{Anisimov1991:PRB}. All that helps to trace the importance and interpretation of the effective spin-orbit couplings. Table \ref{Tab:all-data2} summarizes the relevant spin-orbit couplings including their strengths as taken from Refs.~[\onlinecite{Gmitra2013:PRL,Irmer2015:PRB,Zollner2016:PRB,Frank2016:Cu}]. The general tendency is obvious, the heavier the ad-element, the stronger are the local SOC parameters. Comparing their strengths with respect to the graphene intrinsic SOC, we see that hydrogen and methyl enhance local SOC by two orders of magnitude, fluorine by three orders, and copper enhances local SOC by four orders of magnitude.

There have already been studies constructing model SOC Hamiltonians induced by adatoms in graphene \cite{Weeks2011:PRX,Pachoud2014:PRB}. Our approach to the Hamiltonian building is different from those, so it is not surprising that the forms of the Hamiltonians also differ. The analysis of Weeks~\emph{et al.}~[\onlinecite{Weeks2011:PRX}] focuses on heavy adatoms adsorbed in hollow positions interacting with graphene through three outer $p$-shell orbitals of
the adatom. The fine structure of these orbitals, due to the intra-atomic spin-orbit coupling, gives rise, via hybridization with carbon orbitals, to the induced SOC of the $\pi$ band of graphene. The procedure to integrate out (downfold) the adatom orbitals starts from a fully functionalized graphene with global C$_{6v}$ symmetry---adatoms are occupying each hexagon,---but with no direct coupling between the orbitals on neighboring adatoms.
In contrast, our approach treats a single adatom, so the system has only a local symmetry. Our Hamiltonian for the hollow position thus differs from the one obtained in Ref.~\onlinecite{Weeks2011:PRX}.
In the work of Pachoud~\emph{et al.}~[\onlinecite{Pachoud2014:PRB}], all three relevant adatom configurations are considered, and the choice of the adatom orbitals is not restricted. However, the form of the Hamiltonians is limited to the spatial delta-function (at the adatom site)
multiplied by an $8\times 8$ matrix to cover the pseudospin, valley, and spin spaces. The local structure is thus not preserved, which is not a problem in the continuum limit. Our models instead keep all the local symmetries that adatoms induce (or, rather, still preserve), by assigning pseudospin, spin, and valley-dependent hopping elements in the close neighborhood of the adatom site.

In summary, we have provided in full detail a derivation of effective SOC Hamiltonians for hexagonal systems employing group theory analysis. Our results cover several experimentally relevant scenarios: (i) global SOC effects caused by the proximity of substrates, such as transition metal dichalcogenides, or metallic interfaces; (ii) local SOC effects due to dilute ad-atom or ad-molecule functionalization with emphasis on hollow, top, and bridge adsorption positions.
For both cases (i) and (ii), we have explicitly shown which effective SOC matrix elements are suppressed by the presence or absence of particular symmetries and
classified the SOC mediated hoppings by the subgroups of the full hexagonal point group. In general, our construction-oriented approach is easily transferable to systems with other symmetries and allows one to derive quickly a particular effective SOC Hamiltonian respecting the given symmetries.
Such effective SOC Hamiltonians serve as useful ingredients for model calculations that investigate transport, (quantum) spin Hall effect, spin relaxation and dephasing, WL/WAL measurements, etc.

This work was supported by DFG SFB 689 and GRK 1570, and by the EU Seventh Framework Programme under Grant Agreement No.~604391~Graphene~Flagship.
\bibliography{soc_ped}

\end{document}